\documentclass[sigconf]{acmart}

\usepackage{subfig}
\usepackage{multirow}
\usepackage[]{graphicx}
\usepackage{array, makecell}
\usepackage{amsmath}
\usepackage{xcolor}
\usepackage{enumitem}
\usepackage{booktabs}

\definecolor{LightGreen}{rgb}{0.88,1,0.5}
\definecolor{LightCyan}{rgb}{0.7,1,1}

\AtBeginDocument{%
  \providecommand\BibTeX{{%
    \normalfont B\kern-0.5em{\scshape i\kern-0.25em b}\kern-0.8em\TeX}}}

\copyrightyear{2022} 
\acmYear{2022} 
\setcopyright{acmlicensed}\acmConference[CHI '22]{CHI Conference on Human Factors in Computing Systems}{April 29-May 5, 2022}{New Orleans, LA, USA}
\acmBooktitle{CHI Conference on Human Factors in Computing Systems (CHI '22), April 29-May 5, 2022, New Orleans, LA, USA}
\acmPrice{15.00}
\acmDOI{10.1145/3491102.3517595}
\acmISBN{978-1-4503-9157-3/22/04}

\newcommand{\etals}{{\em et al. }}

\newcommand{\revise}[1]{\textcolor{black}{#1}}

\begin{document}

\title[Does Mode of Digital Contact Tracing Affect User Willingness to Share Information?]{Does Mode of Digital Contact Tracing Affect User Willingness to Share Information? A Quantitative Study}

\author{Camellia Zakaria}
\orcid{0000-0003-4520-9783}
\affiliation{%
  \institution{College of Information \& Computer Sciences, University of Massachusetts Amherst}
  \country{USA}
}

\author{Pin Sym Foong}
\orcid{0000-0002-4437-8326}
\affiliation{%
  \institution{Saw Swee Hock School of Public Health, National University of Singapore}
  \country{Singapore}
}

\author{Chang Siang Lim}
\orcid{0000-0001-9818-6058}
\affiliation{%
  \institution{Future Health Technologies, Singapore ETH Center}
  \country{Singapore}
}

\author{Pavithren V. S. Pakianathan}
\orcid{0000-0002-2232-3658}
\additionalaffiliation{%
  \institution{Singapore University of Technology and Design}
  \country{Singapore}
}
\affiliation{%
  \institution{Saw Swee Hock School of Public Health, National University of Singapore}
  \country{Singapore}
}

\author{Gerald Koh Choon Huat}
\orcid{0000-0002-6453-6897}
\affiliation{%
  \institution{Saw Swee Hock School of Public Health, National University of Singapore}
  \country{Singapore}
}

\author{Simon Tangi Perrault}
\orcid{0000-0002-3105-9350}
\affiliation{%
  \institution{Singapore University of Technology and Design}
  \country{Singapore}
}

\renewcommand{\shortauthors}{Zakaria, et al.}

\begin{abstract}
Digital contact tracing can limit the spread of infectious diseases. Nevertheless, barriers remain to attain sufficient adoption. In this study, we investigate how willingness to participate in contact tracing is affected by two critical factors: the modes of data collection and the type of data collected. We conducted a scenario-based survey study among 220 respondents in the United States (U.S.) to understand their perceptions about contact tracing associated with automated and manual contact tracing methods. The findings indicate a promising use of smartphones and a combination of public health officials and medical health records as information sources. Through a quantitative analysis, we describe how different modalities and individual demographic factors may affect user compliance when participants are asked to provide four key information pieces for contact tracing.
\end{abstract}




\maketitle

\section{Introduction}
\label{sec:intro}
Contact tracing is a public health strategy that plays a crucial role in curbing the spread of communicable diseases \cite{cdc2020}. Manual contact tracing involves a series of steps, starting with contacting infected persons and then interviewing the patient to gather a log of locations and persons with personal contact within a specified period. These close contacts are identified and notified regarding potential exposure and informed of further containment measures such as testing or isolation. However, an incomplete or incorrect recall of events, locations, and contact persons in the period of interest can deter this process \cite{ferretti2020quantifying}. Often, contact tracing is resource-intensive and time consuming \cite{keeling2020efficacy}.

In the U.S., the COVID-19 pandemic has particularly challenged the limits of manual contact tracing. The novel clinical features of COVID-19 infection, including long incubation period, asymptomatic transmission, and high transmission rate \cite{wang2020clinical} led to a burgeoning number of patients that easily outpaced manual contact tracing efforts. These issues are exacerbated by less than 50\% of people being willing to participate in contact tracing efforts \cite{mcclain2020challenges}. 

A modeling study by Feretti \etals \cite{ferretti2020quantifying} shows that a hypothetical, fully automated digital contact tracing solution can slow or stop the transmission of COVID-19. However, the extent of data gathering that digital contact tracing requires leads to concerns with privacy, user surveillance, and data leaks \cite{redmiles2020user,larus2020joint,trust2020pew}, worsened by the lack of trust in government bodies and technology companies to handle sensitive information required in digital contact tracing \cite{cdc2020,sinha2020contact}. These concerns hamper the adoption of digital contact tracing and place more lives at risk. Unfortunately, a recent scoping review concludes that ``there is a dearth of evidence regarding the barriers and facilitators to uptake and engagement with COVID-19 digital contact tracing applications'' \cite{thorneloe2020scoping}. Thus it is essential to determine the factors that increase acceptance of digital contact tracing.

Existing studies prospectively probed the desired feature set to influence users' willingness to participate in digital contact tracing. Researchers have begun to identify a few key facilitators. These are:  higher perceived public health benefit, perceived individual benefit, and lower degree of privacy risk \cite{kaptchuk2020good,megnin2020facilitators}. Missing in this body of knowledge about the users' perception of digital contact tracing features is whether the stated \emph{mode of digital contact tracing} may affect willingness to participate. The perception that a particular mode of delivery is more or less threatening to personal privacy may derail digital contact tracing efforts even before actual usage. While the particular features of an application affect a person's willingness to continue using an application, resistance to the mode of delivery can be a barrier to the \emph{first usage}. To our knowledge, this question of the users’ lens of digital contact tracing modes has not yet been studied.

In this paper, we sought to \textbf{understand how the mode of digital contact tracing affects users' willingness to share different types of data}. We conducted a scenario-based survey of U.S. respondents to understand users' relative willingness to share private information to help contact tracing and examine the modality of data sharing that would be most acceptable by users. We studied three types of information best collected by digital devices to support contact tracing. They are (1) the subject's identity (name and social security number), (2) the subject's contact details (email and contact number), and (3) details of exposure (location or person of interest). We also offered four types of modalities for participants to indicate their willingness to share this information. They are (1) in-person communication with public health officials, (2) providing access to their existing health records, (3) sharing their information collected in their smartphone device (4) providing access to analyze their internet activity. We also examined how such willingness was influenced by demographic variables such as age, parenthood, income, and trust in public health officials. As the study was conducted in the U.S. between November and December 2020, we wanted to ensure that users' responses were not solely based on their experience of the COVID-19 pandemic. Hence, we randomly assigned participants to a different potential epidemic scenario involving one of six infectious agents.

Through a quantitative analysis, we present an overview of users' willingness to share different types of private information to support contact tracing and factors that influence such willingness, supplemented with qualitative responses to help unpack our observations. We found no evidence of the effects of disease types on users' willingness to share their private information for digital contact tracing. Notably, our findings show that participants are most willing to share their private information via smartphones (\emph{$p<.01$}) and grant public health officials access to individuals' health records (\emph{$p<.01$}). However, sharing \emph{location} information is significantly impacted by the level of trust in public health officials (\emph{$p<.01$}). These results add new dimensions to the well-established importance of trust in public health officials in the data collection pipeline for digital contact tracing.

Our work contributes 1) an empirical study of user willingness to adopt contact tracing strategies across various modes of collection, information types, and disease types, and 2) based on these findings, recommendations toward developing a national digital contact tracing strategy in the U.S. 
\section{Related Work}
\label{sec:related}
Our focus here was to summarize existing literature on barriers and facilitators to digital contact tracing. Specifically, we highlight prior work that examined barriers and facilitators of digital contact tracing and the importance of data collection modality.

A recent survey of Americans indicates that just 42\% are willing to download and use a contact-tracing app \cite{zhang2020americans}. For automated and partly automated contact tracing to be effective, at least 56\% uptake is needed \cite{braithwaite2020automated} within a population. What would the barriers and facilitators of digital contact tracing be for sufficient mass adoption in the U.S.?

\subsection{Background on Contact Tracing}
Contact tracing is an essential strategy in public health management, proving itself effective in limiting the spread of infectious disease \cite{krista_swanson_2018}. This process, however, must be performed expeditiously. As defined by CDC, contact tracing involves trained case investigators by the public health organization to evaluate a patient's close contacts (i.e., person-of-interest, POI) \cite{centers2021training}. A POI may expect a case investigator to seek information on their health symptoms and possible exposures (i.e., places, partners) to determine compromised entities during the interview. POI may also receive instructions for isolation and follow-up sessions on their health. Table \ref{tab:contact_tracing_interview} summarises the types of critical, but sensitive information that serves as a general guideline by CDC in conducting such interviews \cite{cdc2020_contactracingguide}. This laborious process can be slow \cite{mirjam2020} and often relies on the POI's willingness and ability to recall the relevant information \cite{peterman1997}. The race against time is critically challenged by the shortage of human resources during a health crisis. Hence, more recently, digital tools were explored to play a role in enhancing contact tracing by either automating or semi-automating some of these data collection tasks.

\begin{table}[h]%\scalebox{1}{
 \caption{A list of open-ended questions for contact tracing by the CDC, and grouped according to information type.}
 \label{tab:contact_tracing_interview}
\resizebox{.9\linewidth}{!}{\begin{tabular}{|l|p{130pt}|p{80pt}|}
\hline
   & \textbf{Questions} & \textbf{Information Type} \\
\hline
1  & What is their name?                                 & \multirow{5}{*}{Identity}\\ \cline{1-2}
2  & What name do they go by?                            & \\ \cline{1-2}
3  & What is their gender?                               & \\ \cline{1-2}
4  & What is their race/ethnicity                        & \\ \cline{1-2}
5  & What is their primary language?                     & \\ \hline
6  & What is the best way to reach them?                 & \multirow{2}{*}{Contact} \\ \cline{1-2}
7  & What is their cell number, email,\par APP and username? & \\ \hline
8  & Where do they live?                                 & \multirow{4}{*}{Exposure Location}                    \\ \cline{1-2}
9  & Who do they live with?                              & \\ \cline{1-2}
10 & Where do they work?                                 & \\ \cline{1-2}
11 & Where is that located?                              & \\ \hline
12 & When did you see them last?                         & Exposure Location\par / Intimate Partner \\ \cline{1-2}
13 & For how long have you spent\par time with them? & \\ \hline
14 & What symptoms might have had? & \multirow{3}{*}{Health Status} \\ \cline{1-2}
15 & What underlying medical\par conditions might they have?& \\ \cline{1-2}
16 & What do they know about your\par  infectious status? & \\ \hline
\end{tabular}}
%}
\end{table}

Based on this established process, our study examines how answers to these questions can be collected through different data collection modalities and not restricted to human contact tracers alone \cite{cdc2020_contactracingguide}.

\subsection{Barriers and Facilitators of Digital Contact Tracing}
Various barriers and facilitators to digital contact tracing have been identified. A discrete choice experiment prospective study in the Netherlands suggested less modifiable individual factors specific to COVID-19 \cite{jonker2020covid}. In this study, the factors that correlated with predicted adoption rates were educational attainment, underlying health conditions, and a perceived threat from COVID-19.

More promising, modifiable facilitators can be gleaned with a broader literature review that includes other diseases. Megnin \etals \cite{megnin2020facilitators} conducted a rapid review of qualitative studies on the factors influencing user uptake and engagement with any contact tracing system across various infectious diseases. They identified four modifiable factors that arise from the users' perception of the application. They are a perceived sense of collective responsibility, perceived personal benefit, the presence of community co-production of contact tracing systems, and the perceived capability of reaching contact persons efficiently and effectively. The authors also identified privacy concerns as a key barrier. These concerns were: mistrust with the requester, unmet needs for information and support, fear of stigmatization (due to an identified infection), and what they called ``mode-specific challenges''. These mode-specific challenges were not communication channel issues per se. They were more akin to accessibility issues (e.g., no smartphone) or usability issues (e.g., difficulty downloading, using the application, or lack of technical proficiency).   

These studies suggest two broader research areas that can be brought to bear upon our knowledge about facilitators and barriers of digital contact tracing. The first area covers the perception of trust in public bodies running the digital contact tracing operations. The second area covers studies on multiple alternative channels of user participation and data collection modes.

\subsection{Issues of Trust and Data Collection}
\label{sec:relatedTrust}

One commonly cited factor that predicts the adoption of contact tracing is the trust in the entity conducting digital contact tracing (i.e., institutional trust). Mayer's body of work on institutional trust defines it as the following: `the willingness of a party to be vulnerable to the actions of another party' \cite{mayer1995integrative}. \revise{In a pandemic, the willingness of users to be vulnerable must be particularly directed towards public health officials, who are providing practical guidelines and general awareness of the disease outbreak. Instead, trust is often referred to as being vulnerable to government bodies and private firms handling the sensitive information required in digital contact tracing.}

\revise{To overcome the mountain of skepticism towards utilizing contact tracing apps,} several public health organizations and researchers \cite{larus2020joint,world2020ethical,kahn2020digital,centers2020preliminary} have proposed that \revise{these} systems be transparent and open to public scrutiny. Scientists and researchers across the globe have also recommended that a privacy-by-design approach be adopted whereby only necessary data is collected and stored using secure encryption techniques to preserve the security and privacy of the data \cite{larus2020joint}. Additionally, the World Health Organization (WHO) and American Civil Liberties Union (ACLU) collectively agree with the CDC's suggestion to make contact tracing voluntary, with full user control over data management \cite{world2020ethical,gillmor2020principles,centers2020preliminary}.

\revise{In response to these recommendations, Google and Apple pushed for anonymous contact tracing \cite{applegoogleCOVID}. The exposure notification framework frequently exchanges anonymous identifier beacons through Bluetooth between and among smartphones whose users are in close proximity; either by installing an official app from their region’s government or directly through a verification protocol from public health authorities. NOVID \cite{novid}, ICheckedIn \cite{apnewsanon} and SaferMe \cite{saferme} are similar efforts that followed in pursuit of anonymity with an encrypted framework. For example, users' names and numbers are hashed in a "pin," maintaining relative obscurity in businesses and places of visit \cite{apnewsanon}. While this information is set to expire in limited duration (e.g., discarded after 30 days), personal information remains accessible to relevant government bodies or pre-approved systems. As might be expected, these efforts did not alleviate the tension associated with the lack of confidence in government and private companies \cite{world2020ethical,cdc2020,altmann2020acceptability,wirth2020citizen}.}

In a study specific to the U.S., only 37\% of respondents found it acceptable to share data with state and local officials, compared to 75\% of the sample who preferred sharing data with infectious diseases researchers \cite{mcclain2020challenges}. Yet, the National Academy for State Health Policy found that contact tracing efforts are either led by state or county (e.g., eight states, including California, are county-led) \cite{nashp2020}.

The suggestion to empower user autonomy in digital contact tracing places the onus on the users to be sufficiently convinced that the benefits of digital contact tracing outweigh the concerns. In a qualitative study on privacy concerns of a tool for online data collection, Phelan \etals \cite{phelan2016s} drew on dual process theories to describe two kinds of user concern about privacy: \emph{intuitive concern} (when following a gut feeling) and \emph{considered concern} (when weighing pros and cons). In \emph{considered concern}, users regularly recognize the benefits of accepting a privacy intrusion, particularly when there is trust in the requester or low assessed risk. However, \emph{intuitive concern} can override the more cognitive path of \emph{considered concern}. This can happen, for example, when the social presence of the requester provides privacy assurance. Additionally, a study on user responses to surveillance suggests that when these types of concerns are not addressed, individuals may adopt protective coping strategies \cite{sanches2020under}. These avoidance mechanisms would make it harder to involve them in contact tracing efforts.

Overall the literature on trust indicates issues of trust seem to be centered on \emph {perceptions} of intrusiveness of the tool. \revise{Although research in anonymous contact tracing actively pushes for computational and technological protocols that strive to guarantee user anonymity and tool credibility, it remains unclear how willing users are to share critical information that must fundamentally be disclosed to public health authorities upon identifying themselves as infected.} Thus, it led us to query if the users' perception of the data collection modality may also affect this intuitive concern. Such knowledge could help pave the way to reducing resistance to privacy intrusions of digital contact tracing for public health.

\subsection{Issues of Technology Access}
\label{sec:issueTechAccess}
Much research on developing the functions of digital contact tracing is centered on the types of sensing modalities available on a smartphone device. Specifically, investigations have looked at Bluetooth and Global Positioning System (GPS) as two standard sensing techniques utilized by most contact tracing mobile applications. Other investigations for collecting location-based data include using a magnetometer, QR code, phone logs, and the most recent WiFi logs of smartphones. \cite{wirth2020citizen,zakaria2022analyzing,trivedi2021wifitrace}. 

However, the primary dependence on a smartphone device for data collection may have the disadvantage of disproportionately disenfranchising several demographic groups. In the U.S., the Pew Research Center reported 66\% smartphone ownership among users aged 65 and above, and 71\% ownership among lower-income earners (less than \$30,000) \cite{sheet2020pew}. In response to a similar situation of low smartphone ownership and complaints about data and power usage in some communities, Singapore's digital contact tracing efforts were supplemented by a dedicated hardware-based Bluetooth token. This self-contained token was distributed in pendant form to the entire population, mainly benefiting people who have reservations about using their smartphone for contact tracing or do not own one. The rollout was complex as it additionally required the distribution of scanning hardware \cite{ttogether} at an estimated cost of about USD 4.6 million \cite{yuen_parliament_2020}. Hence different digital contact tracing modalities present challenges to balance accessibility and privacy \cite{wirth2020citizen, redmiles2020user, centers2020preliminary, gillmor2020principles}.

Given the cost and complexity of digital contact tracing initiatives, it is important to understand how users perceive the modality being used. By selecting modalities that improve the likelihood of uptake, we may increase the chances of success for a digital contact tracing program. 

\subsection {Summary}
Researchers have identified a series of individual user factors that promote the uptake and engagement with a digital contact tracing program. Additionally, the lack of trust in the entities running a digital contact tracing program can negatively influence the acceptability. Finally, the choice of mode should take into account practical concerns of cost and utility for the user and the supplier. Given these considerations, we propose to complement previous work by quantitatively studying how users' perceptions of the mode of digital contact tracing influences users' willingness to share digital contact tracing data. We regard user identification as critical in contact tracing for enforced containment and not just exposure alert (in anonymous tracing); thus, we explicitly consider disclosing personal data following CDC guidelines despite privacy concerns. We also propose to understand the relative acceptability of the different modalities within a spectrum of manual to automated data collection modes.
\section{Data \& Methodology}
\label{sec:dataMethodology}
We conducted an online scenario-based survey study utilizing Amazon Mechanical Turk (MTurk) to understand how previously identified individual characteristics and modes might affect users' willingness to share contact tracing information in different disease outbreak scenarios. Note: MTurk is a crowdsourcing marketplace hosted by Amazon, allowing us to recruit participants (also known as `workers') for our survey research. Prior studies in HCI research have leveraged MTurk to provide researchers with a diverse sample of participants from tens \cite{sannon2019privacy} to hundreds \cite{bentley2016thought}, and thousands \cite{uzor2021investigating}, based on tasks complexity. Later in Section \ref{sec:participantRecruitment}, we provide results from conducting a power analysis to test the probability of our design succeeding with the number of samples acquired from our study.

By not limiting the study to users' responses to one specific disease, our goal was to obtain generalizable knowledge regarding contact tracing for infectious diseases and how they are affected by individual characteristics and data collection modalities. Our study was approved by the university's Institutional Review Board (IRB) and took place between November and December of 2020.

\subsection {Choice of Scenarios}
We selected scenarios from the World Health Organization's (WHO) list of top ten threats to global health, where these communicable diseases account for almost one-third of deaths worldwide \cite{michaud2009global}. The selection of infectious agents covers a broad range of transmission features and participants' level of familiarity given the outbreak histories in the U.S. \cite{michaud2009global}. Diseases were categorized by their transmission methods (i.e., fomite/surfaces, animal vector, air/droplets, sexual transmission). We analyzed them by infectivity and curated a mix of novelty diseases. Based on these criteria, we narrowed the scenario list to six diseases -- Human Immunodeficiency Virus (HIV), Novel Coronavirus (nCov), Zika Virus, Ebola Virus, Methicillin-resistant Staphylococcus Aureus (MRSA), and Hepatitis causing virus (Hep). Table \ref{tab:diseases} shows the feature spread for this selection of diseases.

\begin{table}[h]
\caption{Disease feature spread selected for the scenarios.}
\label{tab:diseases}
\resizebox{\linewidth}{!}{\begin{tabular}{|c|p{55pt}|p{60pt}|p{70pt}|p{67pt}|}
\toprule
\textbf{Disease}\par \textbf{Name} & \textbf{Trans}\par \textbf{mission} & \textbf{Infect}\par \textbf{iousness} & \textbf{Detrimental}\par \textbf{Effects} & \textbf{Comment(s)} \\
\toprule
HIV & Sexual contact & Varies by health\par care context & High lifetime cost & Need sexual \par partner data for\par contact tracing\\ \hline

nCov & Air/Droplets & High & Age and health\par  profile dependent & Current pandemic \\ \hline

Zika & Vector\par (Mosquitoes) & Moderate & 
Specifically to \par  pregnant mothers \par  and fathers-to-be & - \\ \hline

Ebola & Direct Contact,\par bodily liquids & High & Highly fatal & Small U.S. out\par -break in 2014 \\ \hline

MRSA & Surfaces & High & High, Currently\par no Treatment & May occur in\par hospitals \\ \hline

Hep & Sexual contact,\par bodily liquids & Moderate & High & Novel virus but\par older vaccine hints\par  to possible new\par variant outbreak \\ \hline
\end{tabular}}
\end{table}

\subsection{Scenario-based Survey}
\label{sec:userstudy}
We chose to identify each disease in the given scenarios. We considered white-labeling (not naming) each disease scenario to reduce users' existing perceptions of the particular disease. However, after discussion, we realized that presenting the specific characteristics of the disease would still cause participants to recall their understanding or experiences of that characteristic. In contrast, naming the actual disease would improve the likelihood that the responses are real-world perceptions of the disease and the scenarios are NOT associated with COVID-19, which would have had high saliency during the study period.

\begin{figure*}[t]
    \centering
    \subfloat[HIV Scenario.]{\includegraphics[width=0.49\linewidth]{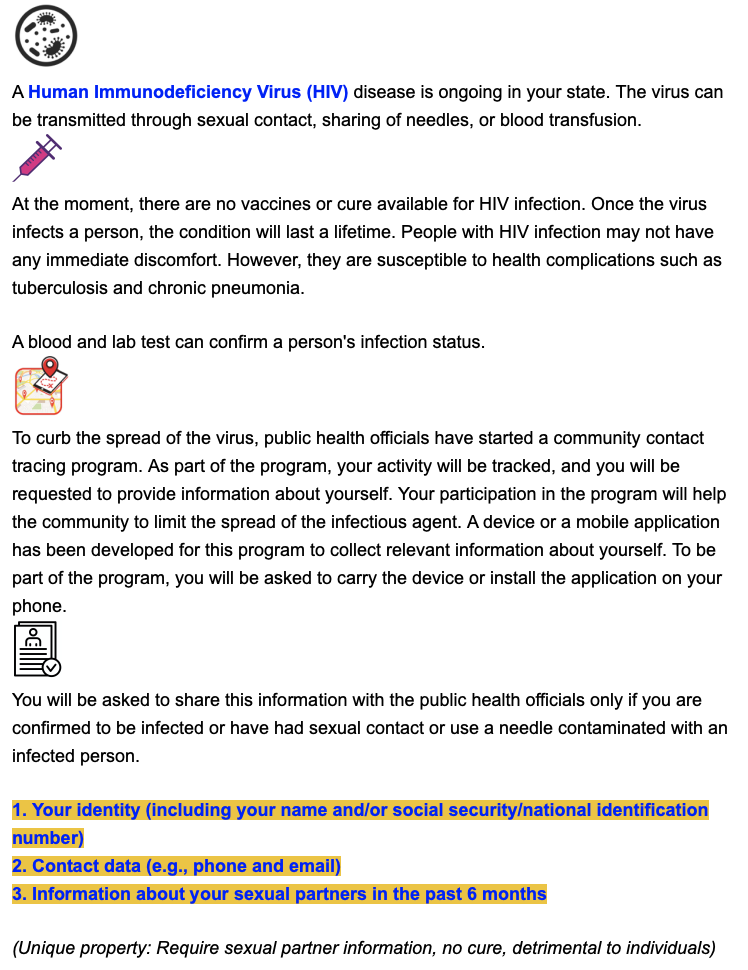}}
    \subfloat[nCov Scenario.]{\includegraphics[width=0.45\linewidth]{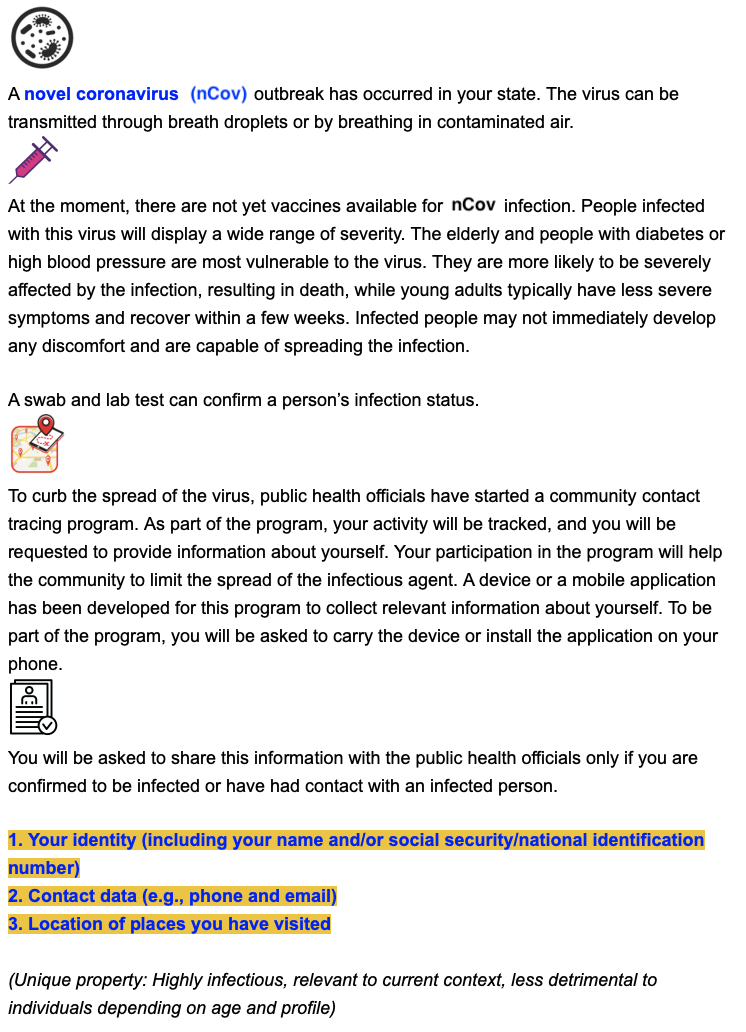}}\vspace*{-6pt}
    \caption{Two of six disease scenario surveys prepared on Qualtrics and integrated with MTurk.}
    \label{fig:scenario}
\end{figure*}

A participant is randomly assigned to a scenario describing one particular type of infectious disease outbreak. As shown in Figure \ref{fig:scenario}, each scenario has the following structure: First, it describes the disease's transmission and severity/lived experience. Then, we provide the current method of confirming infection and, finally, the measures for curbing the disease spread. It is important to note that the reader's/ participant's infection status is not stated in the scenario.

Based on the established contact tracing interview process (see Table \ref{tab:contact_tracing_interview}), we designed each scenario requesting three critical pieces of private information relevant to contact tracing. They are:
\begin{enumerate}%[noitemsep]
    \item Subject's identity defined by their name and social security number,
    \item Subject's contact details defined by their phone number and email address,
    \item Details of exposure are either defined by the locations visited or information of their sexual partners to identify contact transmission.
\end{enumerate}
 
Participants rated their preference on four types of modalities where private information will be collected. These choices represent a spectrum of data collection modes, from an entirely manual collection to a fully automated collection on the user's end. They are: 
\begin{enumerate}%[noitemsep]
    \item Communicating directly with public health officials on the relevant data,
    \item Granting public health officials access to extract the data from existing medical health records,
    \item Providing the data through their smartphone,
    \item Automatically extracting the data from their internet browsing activities.
\end{enumerate}

Our primary consideration for smartphones is with the assumption that the utilization of smartphones is the most common across user groups \cite{sheet2020pew}. We regard a smartphone as central to the usage of other digital device types, including a wearable tracker. Standalone options such as Bluetooth tokens may be used for digital contact tracing \cite{ttogether}. However, the smartphone remains the most immediately viable option for digital contact tracing in the U.S. during the COVID-19 outbreak. Our focus is not to address the usage of possible new modalities.

To understand better user expectations when faced with a disease outbreak event, we asked for a set of multiple-choice questions and open-ended questions. In order of presentation, they are:

\begin{itemize}
	\item A randomly assigned disease scenario as per Figure \ref{fig:scenario}.
	\item Q1: A hygiene question meant to test participants' understanding of the assigned disease scenario.
	\item Q2-5: A set of questions rating participants' willingness to share the three types of private information for digital contact tracing using four types of modality (from 1:Strongly Disagree to 5:Strongly Agree).
	\item Q6: A set of questions asking users to state their most preferred way among the four types of modality for collecting three types of private information for digital contact tracing.
	\item Q7: A set of questions to evaluate participants' trust in public health officials (from 1:Strongly disagree to 5:Strongly Agree). This scale is adapted from McKnight \etals ~\cite{mcknight2002developing} and is used to measure an individual's trusting beliefs towards public health officials across three constructs; competence, benevolence, and integrity. 
	\item Q8-9: A set of questions asking participants about their experience with the disease (0:None, 1: Friends/Family, 2:Personally) and degree of concern with the assigned disease (0:Low, 1:Medium, 2:High). 
	\item Q10-12: A set of open-ended questions querying participants' general concerns with (a) the disease, (b) the selected contact tracing procedure, and (c) their understanding of why that procedure is essential for the disease.
	\item Others: Participants' demographics (age, gender, income, education, marital status, and the number of children. Refer to Appendix \ref{appx:userstudy} for supplementary materials.
\end{itemize}

\subsection {Survey Item Development}
The co-authors of this paper wrote the first drafts of each question, aiming to use phrasing that is brief, relevant, unambiguous, specific, and objective, as recommended by Peterson \etals \cite{peterson_constructing_2000}. The survey questions were then reviewed by an expert panel of two public health researchers, and items were modified based on their feedback. Finally, the questions were prepared in Qualtrics \cite{qual2020}, where they passed the check for readability at 9th-11th grade reading levels.

\subsection{Participant Recruitment}
\label{sec:participantRecruitment}
The survey was then integrated with MTurk, where we selectively defined the qualifications for potential respondents based on the Number of Human Intelligence Tasks (HITs) Approved, HIT Approval Rate, and Location (U.S.). Each participant was randomly assigned to a scenario survey related to one of six disease groups. Each scenario survey was estimated to take 10-15 minutes to complete, with compensation of US\$1.50.

\subsubsection*{Power Analysis}
Our user study was designed to have the same sample size for each of the six disease groups. However, the exclusion of unreliable data unavoidably led to unequal sample sizes. One likely explanation is that we detected plagiarism (high similarity to content searched from the web) within the open-ended responses more frequently in entries that were not related to nCov. This resulted in more exclusion of responses in the non-nCov disease scenarios. However, the relative difference between groups was 4\% or less (Table \ref{tab:summarydemographics}). 

\begin{figure}[h]
\centering
    \includegraphics[width=.8\linewidth]{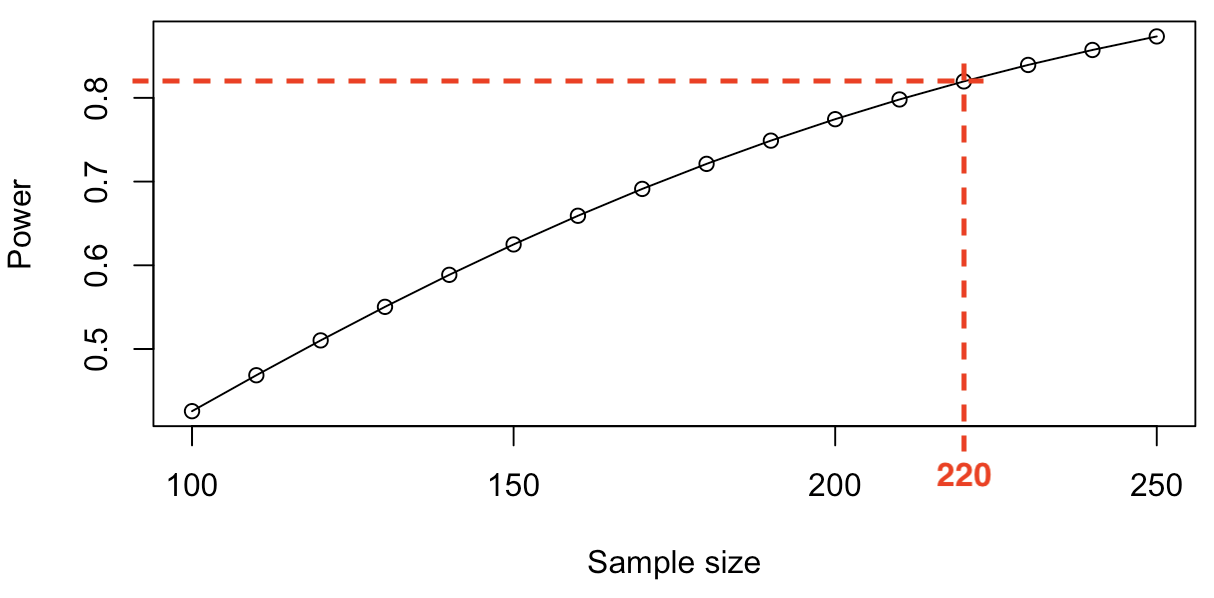}
    \caption{Power curve plots the relationship between sample size and power. A sample size of 220 participants yields a power of 0.8.}
\label{fig:powercurve}
\end{figure}

\begin{table}[h]
    \caption{Power calculation per disease group.}
    \label{tab:powerGroup}
  \begin{tabular}{|l|c|c|}\hline
    \toprule
    \textbf{Disease} & \textbf{n} & \textbf{power} \\
    \toprule
    HIV & 35 & 0.80 \\
    MRSA & 37 & 0.82 \\
    nCov & 43 & 0.89 \\
    Zika & 37 & 0.82 \\
    Hep & 35 & 0.80 \\
    Ebola & 33 & 0.77 \\\hline
    \end{tabular} 
  \end{table}
  
Our analyses aimed to study the relationship between individual factors and user willingness to share information via various contact tracing methods. A power of 0.8 is often desired. The general rule is to assume a medium effect size of 0.25 without information to our experiments \cite{bausell2002power}. Based on our analysis in Figure \ref{fig:powercurve}, our total sample size of 220 would achieve power above 0.8. Table \ref{tab:powerGroup} summarizes the power calculation per disease group. Given our unbalanced design (i.e., the number of samples in each group is not constant), the power for our samples in the Ebola group was slightly under 0.8. However, as will be explained in Section \ref{sec:exploratory}, our experiments moving forward would consider all groups as a whole.

\subsection{Data Collection}
\label{sec:dataMethod}
Table \ref{tab:summarydemographics} summarizes the key details for two rounds of data collection. 

\begin{table}[ht]
 \caption{Summary of 220 verified participants' demographics.}
 \label{tab:summarydemographics}
\centering
\resizebox{\linewidth}{!}{
 \begin{tabular}{p{35pt}p{120pt}|p{35pt}p{100pt}}
 \hline
 \textbf{Parti-}\par \textbf{cipants} & \multicolumn{3}{l}{ 421 total, 314 (verified US National/Resident), 220 (verified entries)} \\
 \hline
 \textbf{Gender} & M (128), F (89), Not stated (3) & \textbf{Period} &  Nov-Dec 2020 \\ \hline
 \textbf{Status} & Married (91), Single (107),\par  Others (22) & \textbf{Parent} & Parent (98), No child (122) \\ \hline
 
 \textbf{Age} & 18-24 (8), 25-34 (94), 35-44 (54), & \textbf{Edu-}\par \textbf{cation} & Skilled (13), Degree (135), \\
 \textbf{} & 45-54 (27), 55-64 (20), > 65 (17) & \textbf{} & Higher Degree (72) \\ \hline
 
 \textbf{Ethn}\par \textbf{-icity} & Asian (12), Caucasian (171), Afr-& \textbf{Region} & Northeast (40), South (96) \\
 \textbf{} & ican American (17), Others (20) &  \textbf{} & West (48), Midwest (36) \\ \hline

 \textbf{Disease} & HIV (35), MRSA (37), nCov (43), & \textbf{Income} & < 25k (102), 25-100k (99), \\
 \textbf{Type} & Zika (37), Hep (35), Ebola (33) & \textbf{} & > 100k (17), NA (2)\\ \hline

 \end{tabular}}
\end{table}

\begin{table*}%[b]
  \caption{Summary statistics in order of survey presentation. $\bar{x}$ denotes sample mean, \emph{M} is sample median, and \emph{s} is the standard deviation of a sample. TRUE refers to participants correctly answering the hygiene question. Willingness to share data, modes for Contact Tracing (CT), and Trust in public health officials scores are on a 1-5 Likert scale. Disease experience scores 0 for no experience, 1 for friends and family, 2 for personal experience, 3 for more than one experience. Disease concern counts participants expressing high, medium and low concern.}
 \label{tab:summ_stats}
\centering
\resizebox{.85\textwidth}{!}{
 \begin{tabular}{|l|c|c|c|} \hline
 \toprule 
 \textbf{Data} & \textbf{HIV} & \textbf{MRSA} & \textbf{nCov} \\
 \toprule
 \textit{Hygiene} & TRUE:30, FALSE:5 & TRUE:37, FALSE:0 & TRUE:42, FALSE:1 \\ \hline
 %\textit{Hygiene} & TRUE:85\%, FALSE:15\% & TRUE:100\%, FALSE:0\% & TRUE:98\%, FALSE:2\% \\ \hline
 \textit{Share location data} & - & $\bar{x}$: 3.55, \emph{M}: 4, \emph{s}: 1.22 & $\bar{x}$: 3, \emph{M}: 3.25, \emph{s}: 1.08 \\
 \textit{Share sex partner data} & $\bar{x}$: 2.43, \emph{M}: 2, \emph{s}: 1.31 & - & - \\
 \textit{Share contact data} & $\bar{x}$: 2.62, \emph{M}: 2.75, \emph{s}: 1.39 & $\bar{x}$: 3.48, \emph{M}: 4, \emph{s}: 1.31 & $\bar{x}$: 2.99, \emph{M}: 3.25, \emph{s}: 1.11 \\
 \textit{Share personal data} & $\bar{x}$: 2.54, \emph{M}: 2.5, \emph{s}: 1.36 & $\bar{x}$: 3.22, \emph{M}: 3.75, \emph{s}: 1.24 & $\bar{x}$: 2.88, \emph{M}: 3, \emph{s}: 1.06 \\\hline
 \textit{CT by public health officials} & $\bar{x}$: 2.63, \emph{M}: 3, \emph{s}: 1.38 & $\bar{x}$: 3.80, \emph{M}: 4, \emph{s}: 1.3 & $\bar{x}$: 3.4, \emph{M}: 4, \emph{s}: 1.21 \\
 \textit{CT through health records} & $\bar{x}$: 2.75, \emph{M}: 2.67, \emph{s}: 1.5 & $\bar{x}$: 3.68, \emph{M}: 4, \emph{s}: 1.36 & $\bar{x}$: 3.22, \emph{M}: 3.67, \emph{s}: 1.3 \\
 \textit{CT through smartphone} & $\bar{x}$: 2.54, \emph{M}: 2.67, \emph{s}: 1.42 & $\bar{x}$: 3.44, \emph{M}: 4, \emph{s}: 1.38 & $\bar{x}$: 2.97, \emph{M}: 3.33, \emph{s}: 1.27 \\
 \textit{CT through internet} & $\bar{x}$: 2.19, \emph{M}: 1.67, \emph{s}: 1.41 & $\bar{x}$: 2.74, \emph{M}: 3.17, \emph{s}: 1.43 & $\bar{x}$: 2.24, \emph{M}: 2, \emph{s}: 1.26 \\ \hline
 \textit{Trust in public health officials} & $\bar{x}$: 3.62, \emph{M}: 3.82, \emph{s}: 0.99 & $\bar{x}$: 3.99, \emph{M}: 4.27, \emph{s}: 1.03 & $\bar{x}$: 3.63, \emph{M}: 3.91, \emph{s}: 0.91 \\ \hline
 \textit{Disease Experience} & 0: 27, 1: 7, 2: 1, 3: 0 & 0: 27, 1: 6, 2: 4, 3: 0 & 0: 19, 1: 18, 2: 3, 3: 3 \\ \hline
 %\textit{Disease Experience} & 0: 77\%, 1: 2\%, 2: 1\%, 3: 0 & 0: 73\%, 1: 16\%, 2: 11\%, 3: 0 & 0: 44\%, 1: 42\%, 2: 7\%, 3: 7\% \\ \hline
 \textit{Disease Concern} & High: 6, Med: 9, Low: 20 & High: 5, Med: 19, Low: 13 & High: 18, Med: 20, Low: 5 \\ \hline
 %\textit{Disease Concern} & High: 17\%, Med: 26\%, Low: 57\% & High: 14\%, Med: 51\%, Low: 35\% & High: 42\%, Med: 47\%, Low: 12\% \\ \hline
 \toprule \textbf{Data} & \textbf{Hep} & \textbf{Zika} & \textbf{Ebola} \\\hline
 \toprule
 \textit{Hygiene} & TRUE:34, FALSE:1 & TRUE:31, FALSE:6 & TRUE:27, FALSE:6 \\ \hline
 \textit{Share location data} & $\bar{x}$: 2.91, \emph{M}: 3, \emph{s}: 1.14 & $\bar{x}$: 2.97, \emph{M}: 3.38, \emph{s}: 1.42 & $\bar{x}$: 3.31, \emph{M}: 3.5, \emph{s}: 1.04 \\
 \textit{Share contact data} & $\bar{x}$: 2.81, \emph{M}: 3, \emph{s}: 1.13 & $\bar{x}$: 2.92, \emph{M}: 3.25, \emph{s}: 1.32 & $\bar{x}$: 3.25, \emph{M}: 3.5, \emph{s}: 1.08 \\
 \textit{Share personal data} & $\bar{x}$: 2.62, \emph{M}: 2.5, \emph{s}: 1.11 & $\bar{x}$: 2.68, \emph{M}: 2.88, \emph{s}: 1.18 & $\bar{x}$: 3.14, \emph{M}: 3.25, \emph{s}: 1.01 \\\hline
 \textit{CT by public health officials} & $\bar{x}$: 3.29, \emph{M}: 3.67, \emph{s}: 1.27 & $\bar{x}$: 3.07, \emph{M}: 3.67, \emph{s}: 1.38 & $\bar{x}$: 3.60, \emph{M}: 4, \emph{s}: 0.99 \\
 \textit{CT through health records} & $\bar{x}$: 3.06, \emph{M}: 3.33, \emph{s}: 1.28 & $\bar{x}$: 3.04, \emph{M}: 3.5, \emph{s}: 1.29 & $\bar{x}$: 3.33, \emph{M}: 3.67, \emph{s}: 1.27 \\
 \textit{CT through smartphone} & $\bar{x}$: 2.77, \emph{M}: 3, \emph{s}: 1.38 & $\bar{x}$: 2.97, \emph{M}: 3.33, \emph{s}: 1.33 & $\bar{x}$: 3.33, \emph{M}: 4, \emph{s}: 1.24 \\
 \textit{CT through internet} & $\bar{x}$: 2.01, \emph{M}: 1.67, \emph{s}: 1.14 & $\bar{x}$: 2.34, \emph{M}: 2.33, \emph{s}: 1.3 & $\bar{x}$: 2.67, \emph{M}: 2.33, \emph{s}: 1.37 \\ \hline
 \textit{Trust in public health officials} & $\bar{x}$: 3.68, \emph{M}: 3.82, \emph{s}: 0.88 & $\bar{x}$: 3.34, \emph{M}: 3.68, \emph{s}: 1.01 & $\bar{x}$: 3.71, \emph{M}: 3.73, \emph{s}: 0.73 \\ \hline
 \textit{Disease Experience} & 0: 26, 1: 8, 2: 1, 3: 0 & 0: 34, 1: 3, 2: 0, 3: 0 & 0: 23, 1: 6, 2: 4, 3: 0 \\ \hline
 \textit{Disease Concern} & High: 5, Med: 9, Low: 21 & High: 2, Med: 10, Low: 25 & High: 8, Med: 8, Low: 17 \\ \hline
 \end{tabular}}
\end{table*}

\subsubsection*{Verification Procedure}
We recruited a total of 421 participants with 314 participants stating their nationality and region of residence in the U.S. Our final analyses relied on verified responses from 220 participants. Verification was performed by our research team manually inspecting the qualitative data answering open-ended questions.

Ninety-four responses were discarded. The removal steps were: First, if a participant supplied an incorrect answer to the hygiene test (Q1), we verified the survey's legitimacy through their qualitative responses (Q10-12). Secondly, we removed entries if any response to the qualitative questions had plagiarized content searched on the web or repeated exact words. 19 out of 220 valid responses provided a wrong answer to the hygiene question. They were still considered valid as their qualitative questions were not plagiarized in any way, suggesting the presence of personal input by the participant. 

\subsubsection*{Coding Procedure}
Thematic analysis \cite{braun_using_2006} was used to analyze the open-ended answers. One researcher generated codes and themes by reviewing all participants' responses, and disagreements on themes were iteratively resolved with the team (refer to Appendix \ref{appx:codebook}). Two researchers applied the pre-established code to our data set and achieved a Cohen's $ \kappa = 0.816 $ (strong level of agreement), where 1 is perfect agreement.

\begin{figure*}%[b]
    \centering
    \subfloat{\includegraphics[width=\linewidth]{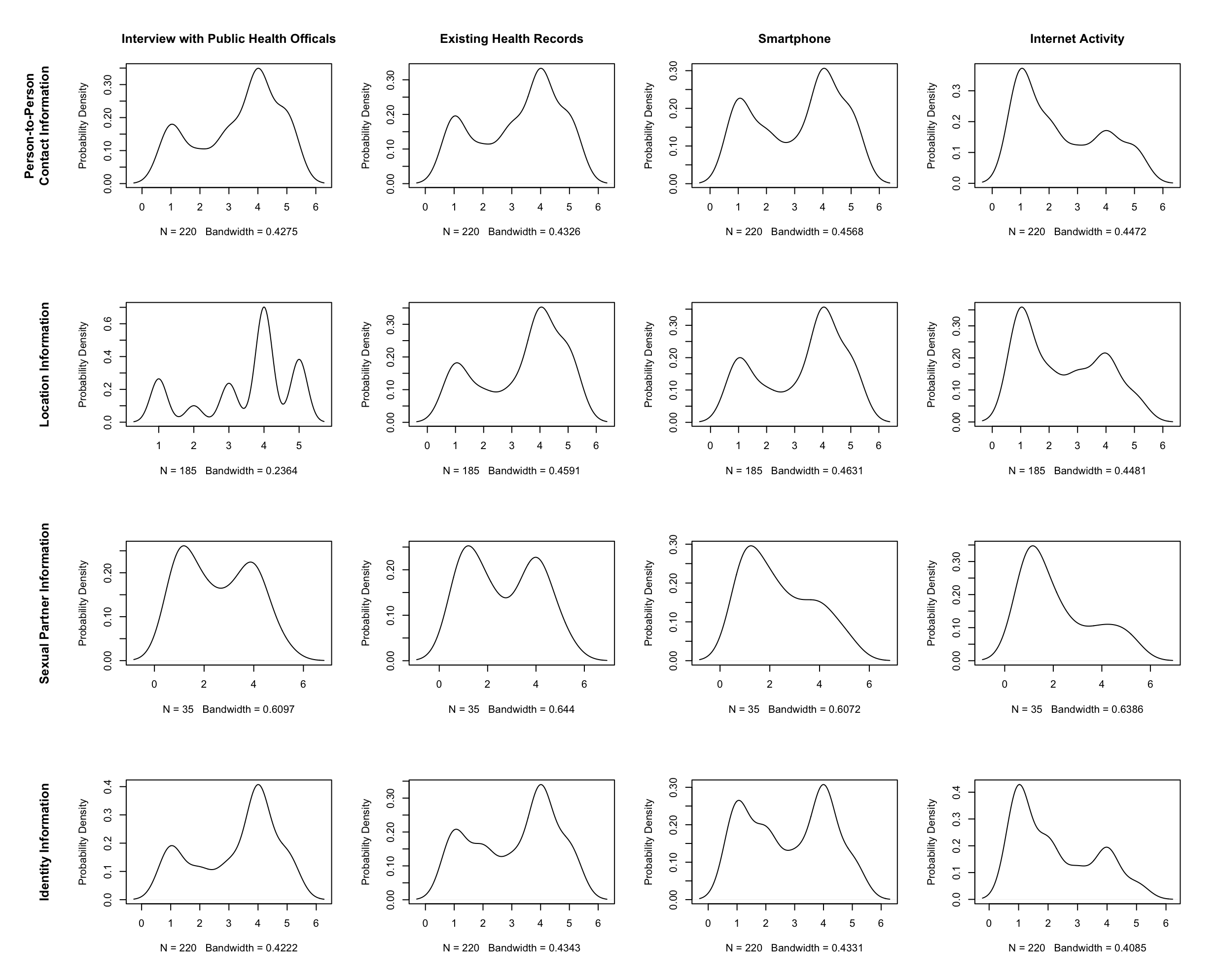}}
    \caption{Probability distribution of willingness to share information through different modality.}
    \label{fig:infoModalities}
\end{figure*}

\section{Exploratory Analysis}
\label{sec:exploratory}
Table \ref{tab:summ_stats} provides a summary of data (N=220) grouped by six disease types in order of survey presentation. Despite applying data transformation, results from conducting a Shapiro-Wilk test for data normality achieved statistical significance ($p<.001$) implying that the distribution of the data are significantly different from normal distribution and thus, we cannot assume normality. We compared the variances of six groups with Levene test, which is less sensitive to departures from normality based on the median. \revise{The test results in no significant differences (location: $p=.119$, contact: $p=.147$, and identity: $p=.127$).} With no evidence suggesting that the variance of information sharing is significantly different for the disease groups, we test the effects of diseases and willingness to share information \revise{(location: $p=.088$, contact: $p=.055$, and identity: $p=.098$)}. The Kruskal-Wallis test revealed no significant difference (\emph{$p>.05$}), thus did not provide strong evidence to reject the null hypothesis that these disease groups have identical means. Separately, post-tests are more conservative to find differences between groups. Since the data are not normally distributed, we employed the Dunn's test with FDR adjustment as a nonparametric post-hoc alternative. The tests yield no statistical difference in means \revise{(\emph{$p>.05$})}, indicating the data can be treated as a whole, except for an HIV scenario which requires a different set of contact tracing information (see Section \ref{sec:userstudy}). Appendix \ref{appx:exploratory} \revise{reports the statistical results for these tests; Figures \ref{tab:locationPlot}, \ref{tab:contactPlot} and \ref{tab:identityPlot} chart the differences between groups, while Tables \ref{tab:locationDunn}, \ref{tab:contactDunn} and \ref{tab:identityDunn} summarize the p-values between two disease groups for every data type.}

%there were tendencies toward differences between the disease states (as depicted in Appendix C) which may compromise the integrity of the strategy to combine them for analysis,
%The test results in indistinguishable differences (\emph{$p>.1$}). 

\begin{figure*}%[b]
    \centering
    \subfloat[Public health officer]{\includegraphics[width=.25\linewidth]{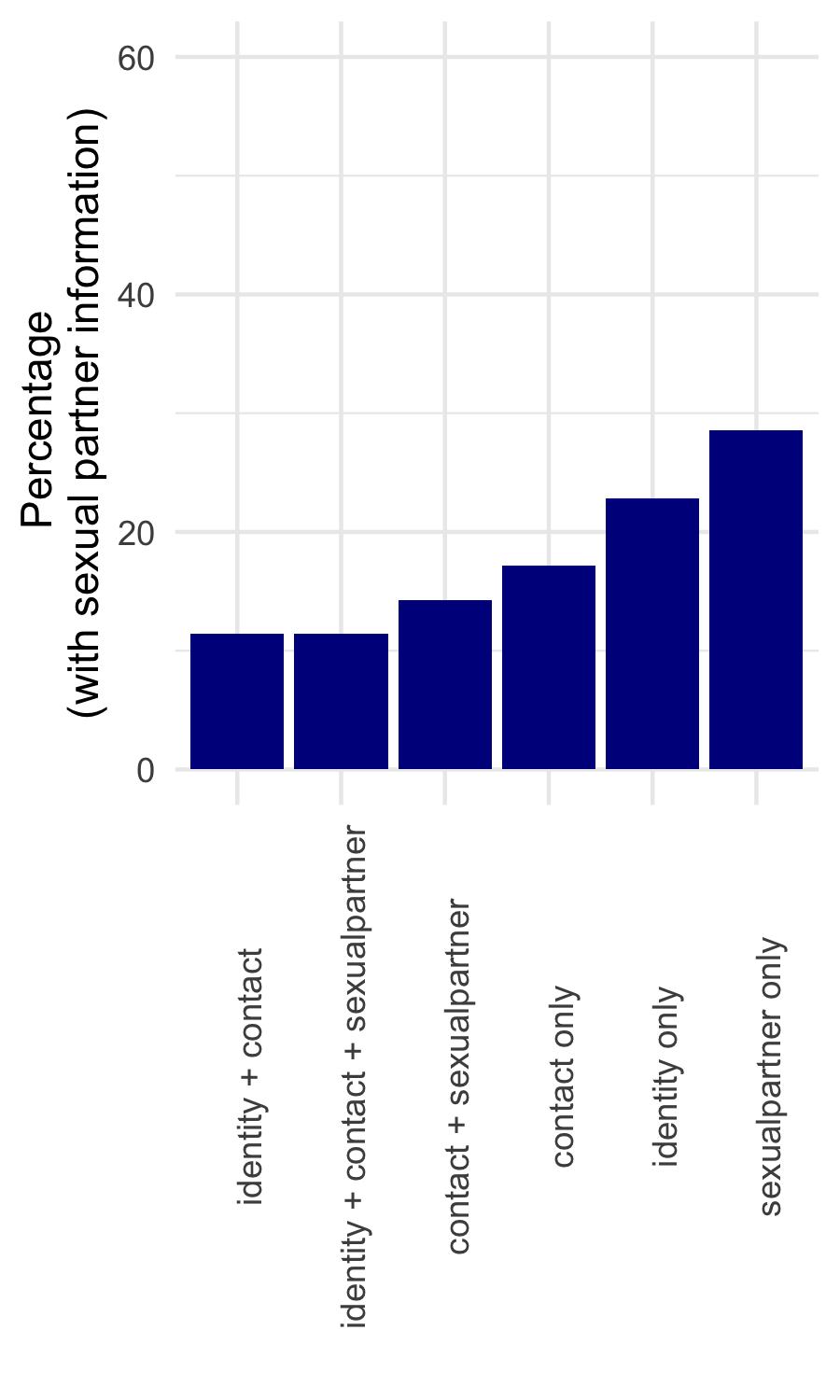}}
    \subfloat[Health records]{\includegraphics[width=.25\linewidth]{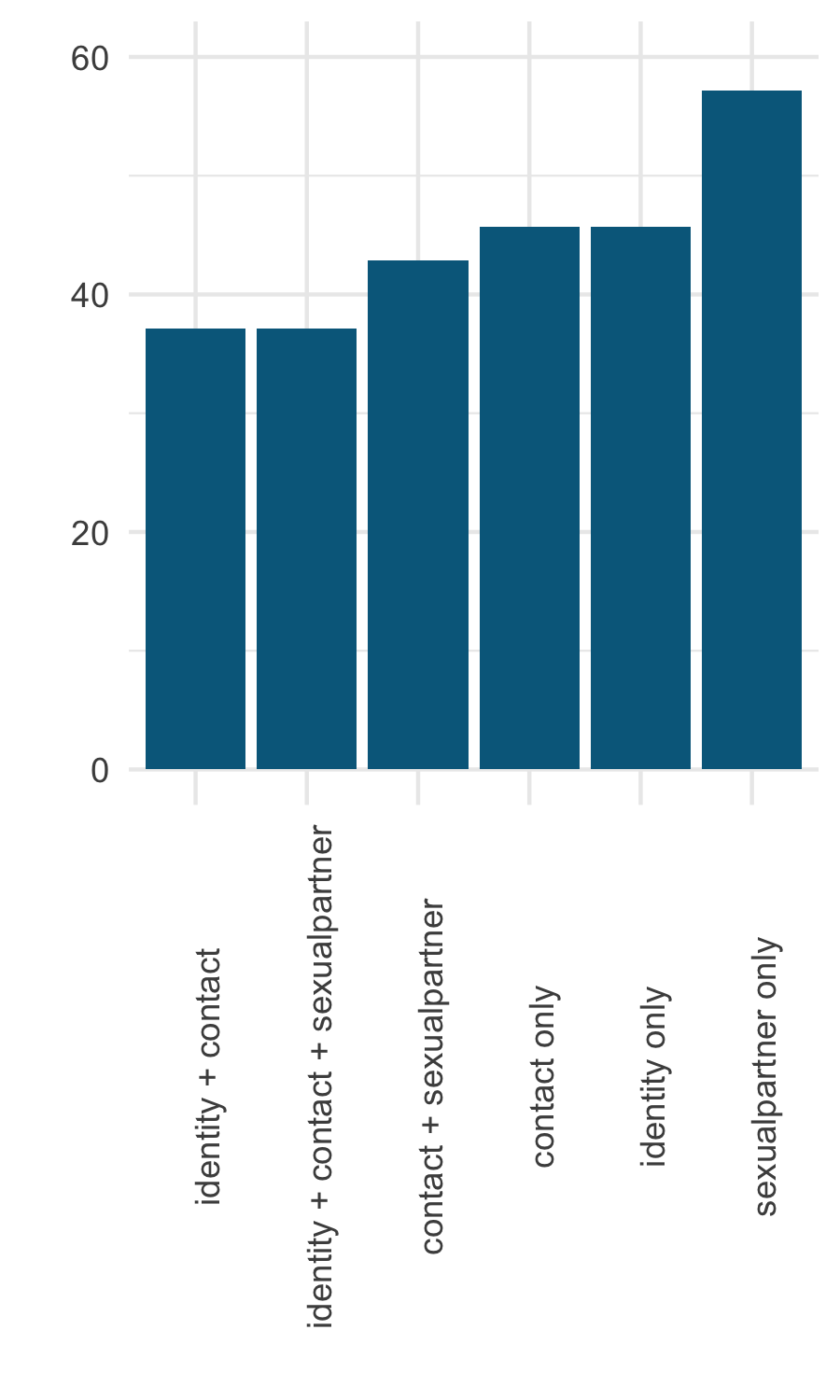}}%\\
    \subfloat[Smartphone]{\includegraphics[width=.25\linewidth]{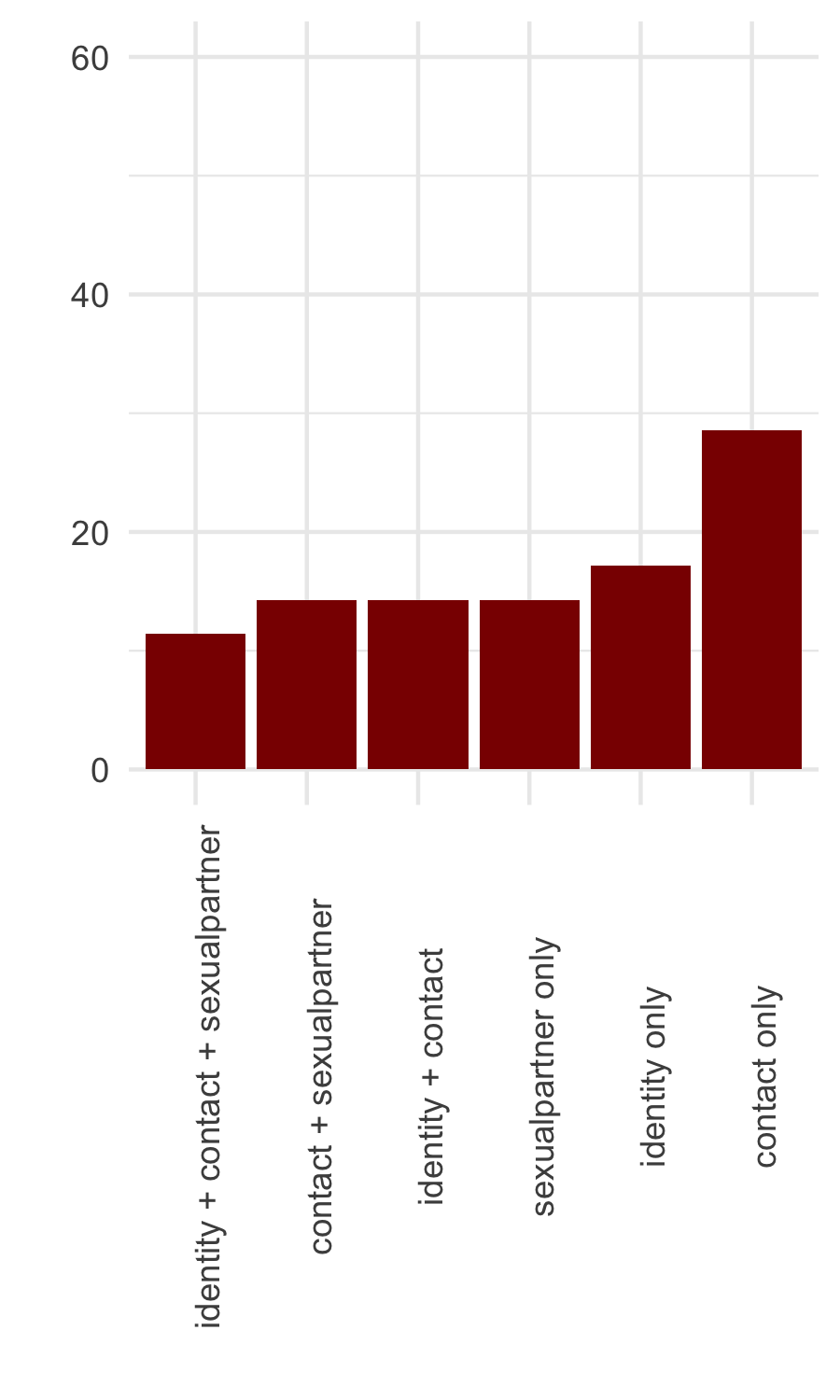}}
    \subfloat[Internet browsing]{\includegraphics[width=.25\linewidth]{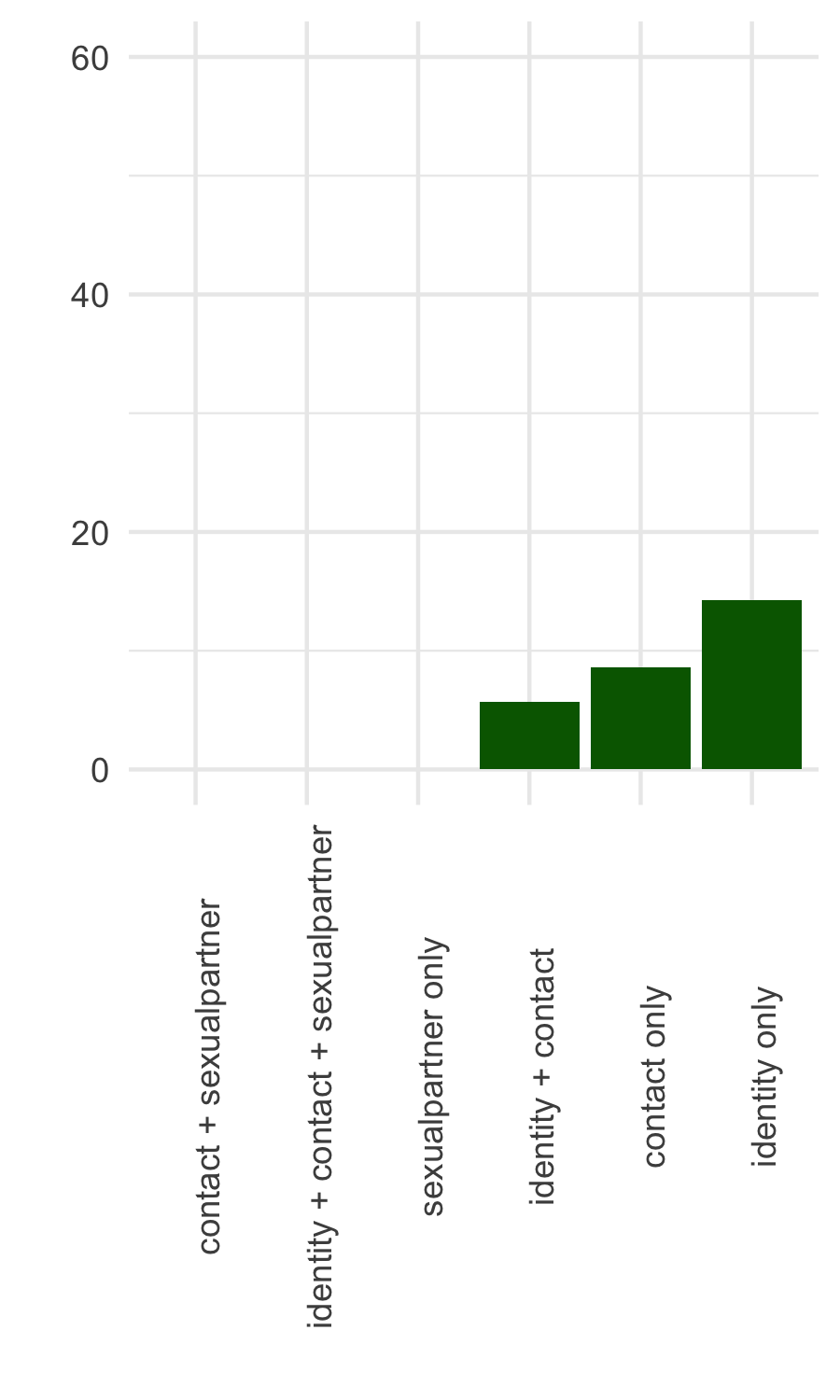}}
    \caption{Percentage of users preference to share data, a combination of personal identifiable, contact details and sexual partner PII for HIV disease, through different methods.}
    \label{fig:infoModalitiesHiv}
\end{figure*}

\begin{figure*}[t]
    \centering
    \subfloat[Public health officer]{\includegraphics[width=.25\linewidth]{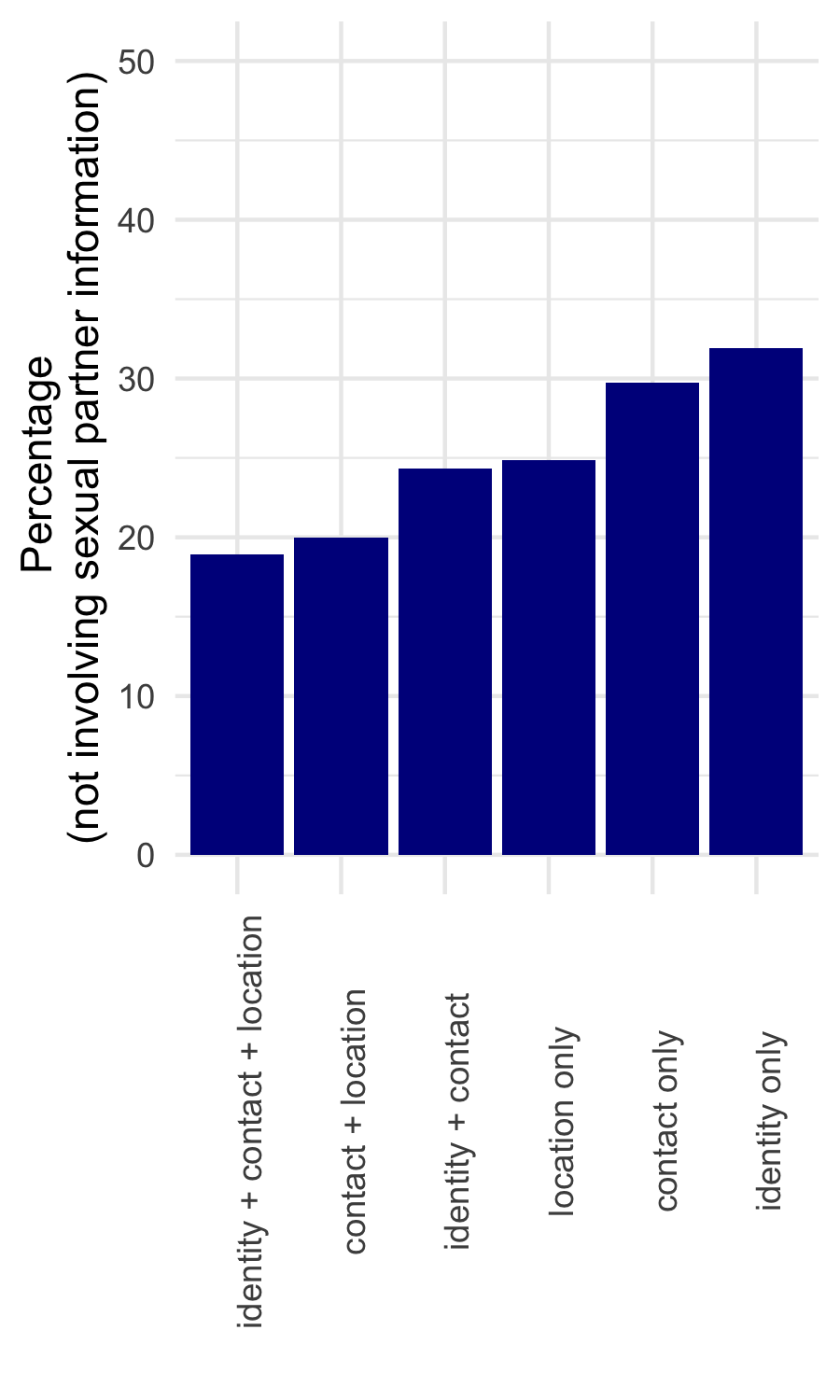}}
    \subfloat[Health records]{\includegraphics[width=.25\linewidth]{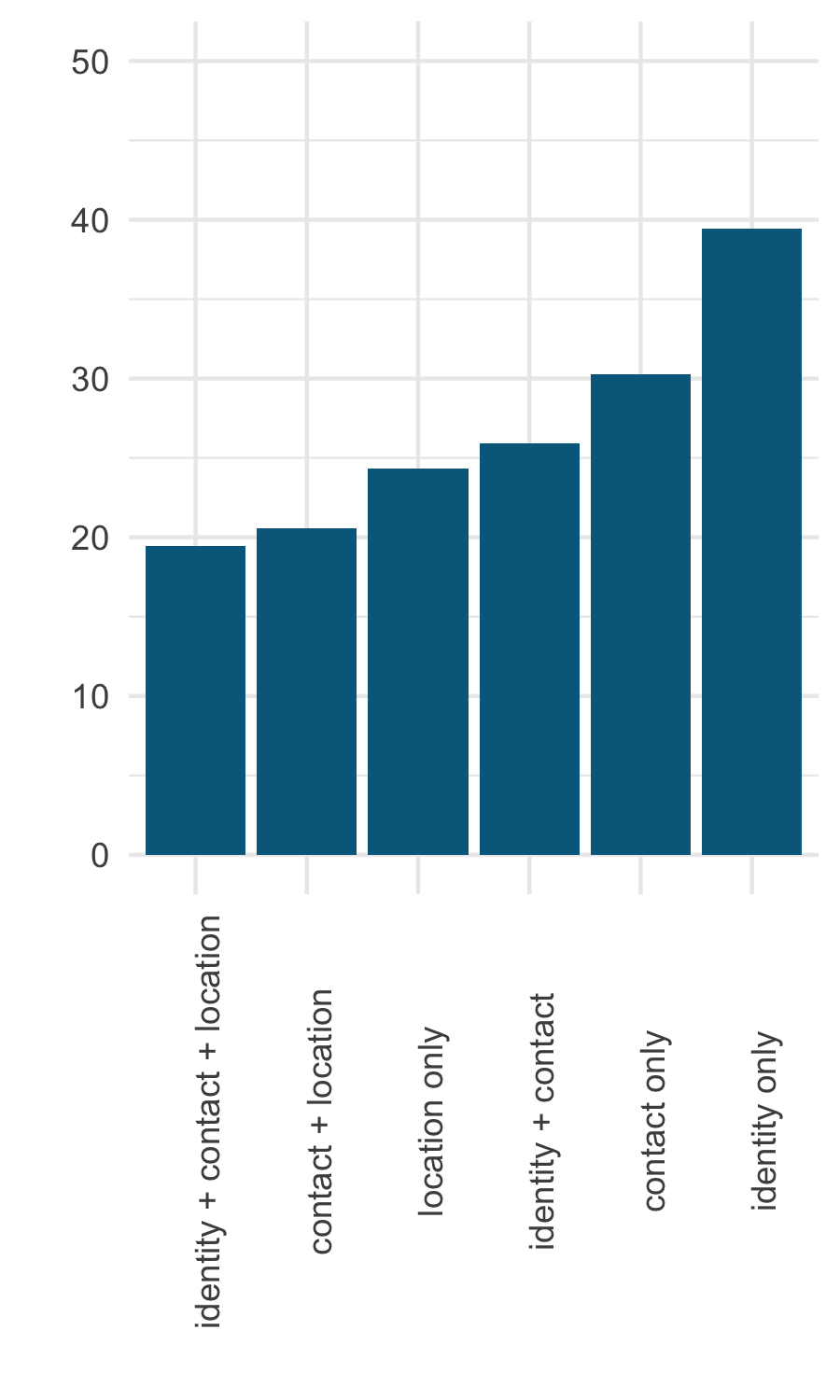}}%\\
    \subfloat[Smartphone]{\includegraphics[width=.25\linewidth]{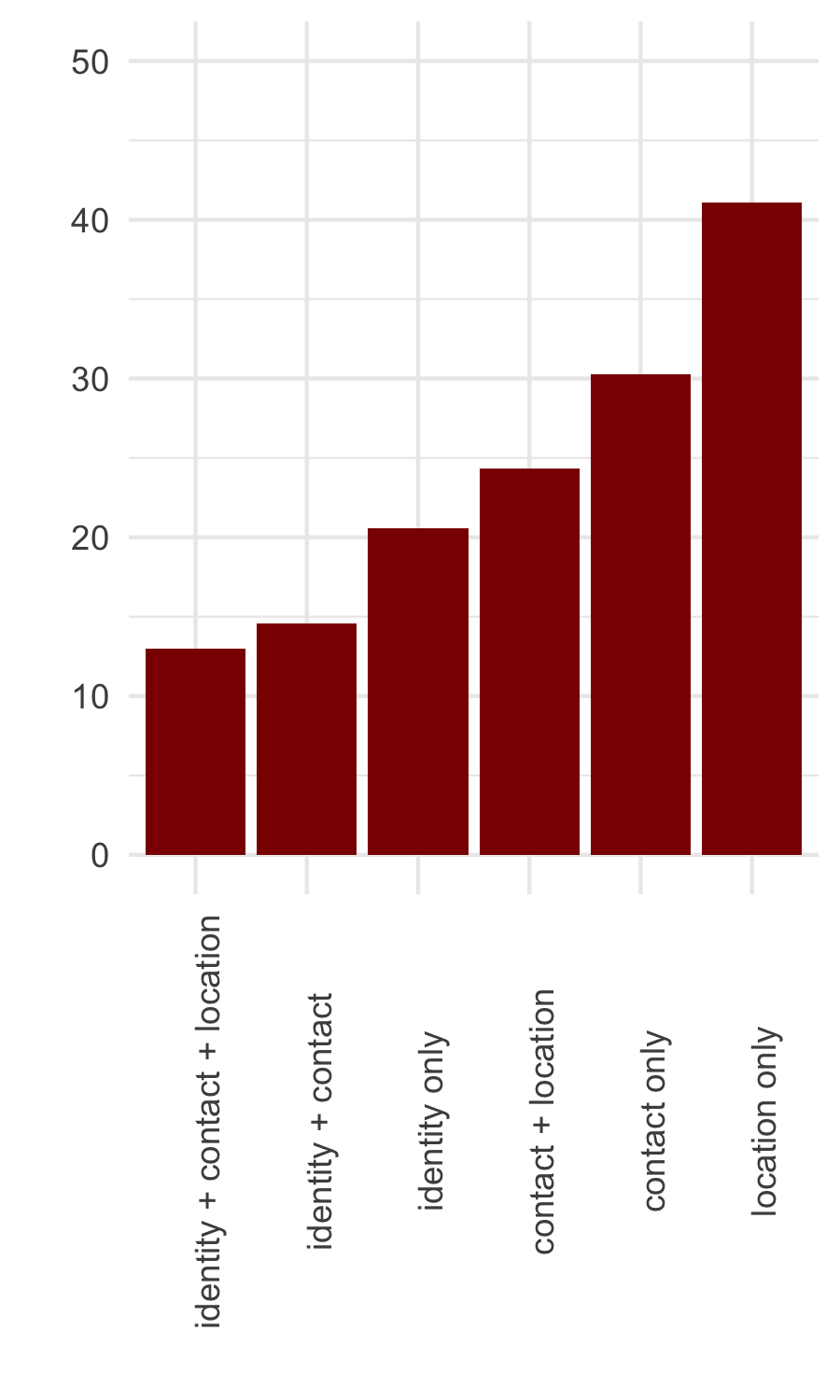}}
    \subfloat[Internet browsing]{\includegraphics[width=.25\linewidth]{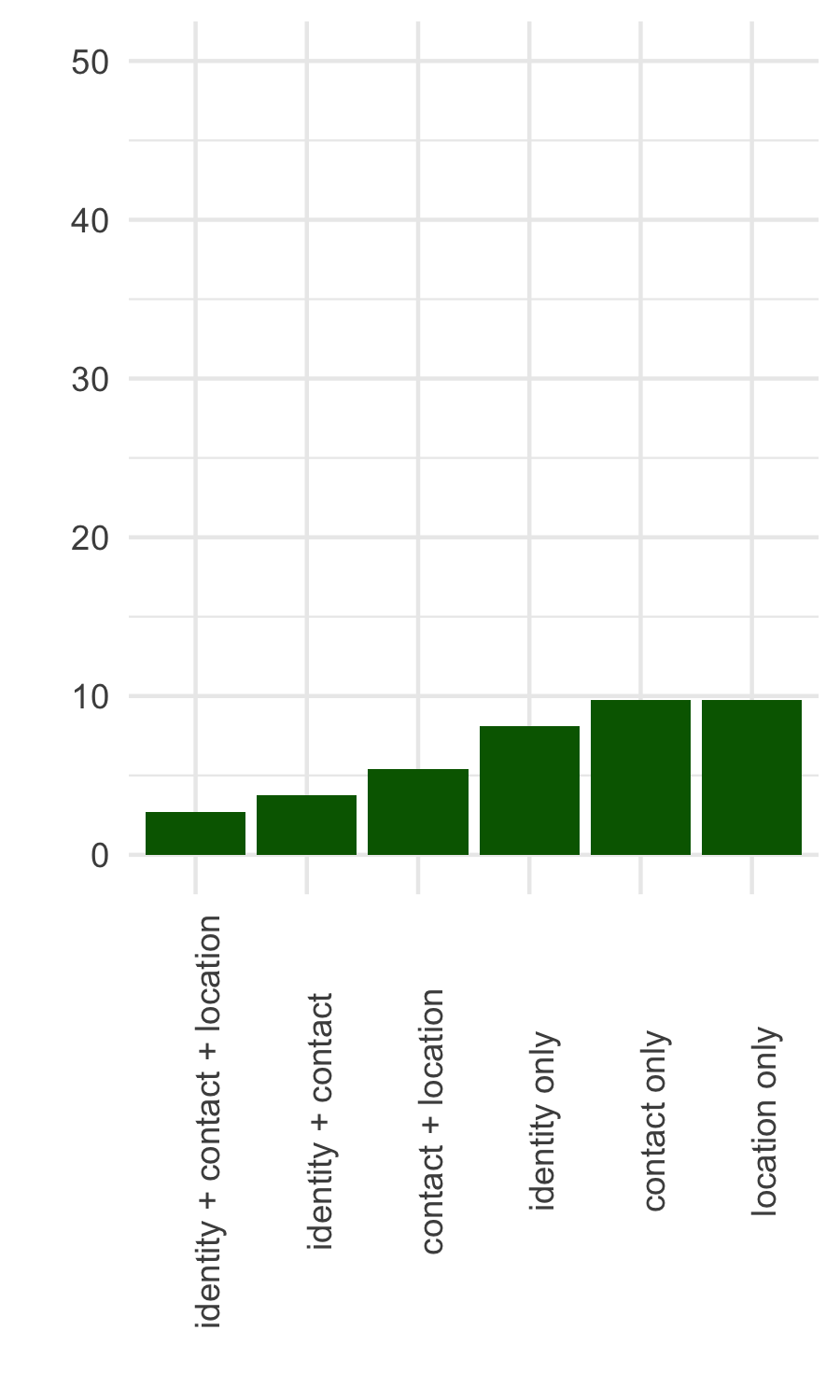}}
    \caption{Percentage of users preference to share data, a combination of personal identifiable information, contact  details and location information for most diseases, through different methods.}
    \label{fig:infoModalitiesAll}
\end{figure*}

Despite the similarity between disease scenarios, in the following paragraphs we point out some data of note. Note: when reporting the results, \emph{M} stands for median and $\bar{x}$ for mean.

Overall, 91\% of our participants indicated a correct understanding of the disease's nature by providing the true answer to the hygiene question (Q1). All participants were asked about their experience with the disease (Q8). All disease groups except nCov had most participants reporting no experience of the disease (Q8). One participant reported personal experience with HIV, while 24 participants noted some personal experience with nCov.

In expressing their concerns on the diseases (Q9), participants expressed the lowest concern for a Zika outbreak (n=25). Most participants expressed deep concern for a nCov outbreak (n=18). These statistics are consistent with the current state of knowledge on disease threats. Specifically, only four Zika transmissions were reported to have occurred in the U.S. since 2020 \cite{cdczika2020}. On the other hand, with the most recent coronavirus first identified in January 2020 (COVID-19) and the country battling over 30 million cases (as of when this study was conducted), it is reasonable to expect the concern for this disease to be high \cite{nytcovid}.

Participants were asked about their trust in public health officials, as defined by the modified ``trusting belief'' construct from \cite{mcknight2002developing} (see Section \ref{sec:userstudy}). Trust in public health officials is moderately high among all participants across disease groups ($\bar{X}$: 3.66/5, \emph{M}: 3.91/5, \emph{SD}: 0.94). This result is consistent with Pew Research's  reporting that a majority of Americans are somewhat confident with public health organizations \cite{trust2020pew}.

Participants expressed more willingness to share private information for tracing MRSA (e.g., location: \emph{M}:4, contact: \emph{M}:4, personal: \emph{M}:3.75) and Ebola (e.g., location: \emph{M}:3.5, contact: \emph{M}:3.5, personal: \emph{M}:3.25) compared to other diseases. It is worth noting that users' willingness to share private information is lowest among our HIV sample, particularly in providing sex partner information. Our results support Njozing \etals findings that highlight the challenges in persuading patients to disclose their HIV status to their partners after testing \cite{njozing2011if}.

While more participants generally preferred communicating directly with a public health officer for contact tracing, scores for other modalities are consistently higher among our MRSA sample (e.g., \emph{M} for contact tracing methods -- smartphone: 4, internet: 3.17, public health officer: 4, health record: 4). From our review of the open-ended responses, it seems that MRSA is perceived as being a high threat. For example, participant P1 responded, ``MRSA is one of the most common antibiotic-resistant bacteria. [It] can occur between food[,] animals and people,'' and P2 described it as ``a very bad situation [to have] a resistance bacterial outbreak.''
On the other end of the scale, participants in the HIV scenario rated the modality preferences lowest. These results are consistent with prior findings on contact tracing pushback in response to HIV, primarily due to the stigma of the virus \cite{davis2020contact}. P3 expressed, ``I think that personal information and your status [should not] be attached to your identity. There is a negative stigma with HIV positive individuals.'' P4 wrote, ``If I was infected, I would be concerned about suffering discrimination from those assuming the worst about how I got infected.'' 

\subsection{Mixing Modalities and Reducing Data Dimensionality}
We plotted Figure \ref{fig:infoModalities} to examine if user willingness to share different types of information can improve through different data collection modalities. We observed an overall decrease in user willingness when the data collection method shifted from manual to increasingly automated methods. Lying on the most extreme of user willingness, sharing any kind of private information through internet browsing histories remains the least acceptable method among users in all scenarios.

Taking a step further, we found that reducing data dimensionality for each modality offered more promising results. As shown in Figures \ref{fig:infoModalitiesHiv} and \ref{fig:infoModalitiesAll}, we observed increasing users' acceptance of a tool when it was not set to collect all three critical types of contact tracing information. Instead, most of our participants preferred using a mix of modalities to share their private information, and these preferences differ with information types. Specifically, most users would prefer to either communicate with a public health officer about their identity and contact information (>30\%) or grant access to their health records of this information ($\approx$40\%). Collecting location information using a smartphone remains favorable. However, user willingness decreased to less than 20\% when identity and contact information are collected alongside location histories.

In the context of an HIV scenario, our results showed a higher preference for utilizing health records to retrieve information related to their sex partners (>50\%). This preference is in contrast to non-sexually transmitted infections, where participants were more positive towards communicating private information directly with a public health official and utilizing a smartphone to share such information.

\subsection{Visualization of Qualitative Responses}
\label{sec:qualAnalysis}
\begin{figure}[h!]
    \centering
    \subfloat[On contracting disease]{\includegraphics[width=.33\linewidth]{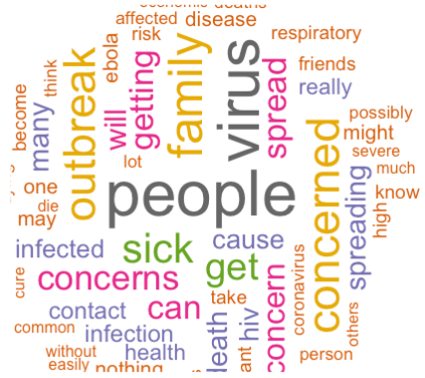}}
    \subfloat[On sharing data]{\includegraphics[width=.33\linewidth]{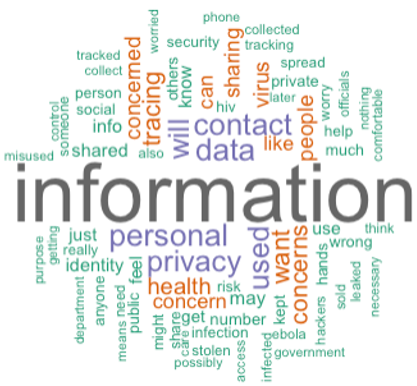}}
    \subfloat[Value of CT]{\includegraphics[width=.33\linewidth]{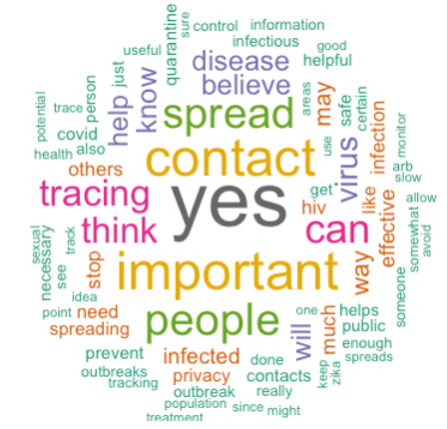}}
    \caption{Summary of the most salient issues captured in users'
     qualitative responses}
    \label{fig:wordcloud}
\end{figure}

As a final step, we generated word clouds to visualize our qualitative responses in Figure \ref{fig:wordcloud}. Participants were asked about their concerns about the disease type, sharing their private information for contact tracing and whether contact tracing is important. It is important to note that while these representations do not capture the meaning of the content, they help identify the most salient issues in this exploratory phase.

Firstly, our users' two most prominent words are `people' and `virus' when asked about their general concerns about the disease. These words are commonly associated with \emph{fear of contracting disease}, followed by \emph{fear of death}. Despite expressing extreme worries, their concerns with sharing contact tracing information for the disease are mainly related to \emph{loss of privacy} and \emph{misuse of data}. Words such as `information,' `personal privacy,' `contact,' and `used' from their expressed concerns conform with our understanding of why contact tracing faced strong community resistance during the COVID-19 pandemic \cite{sinha2020contact}. Some participants expressed worry that their data would fall into the hands of the wrong organization. P5 reported, ``I [am] concerned about [a] data breach that we hear about with other companies having our data stolen and lost. I [do not] want identity theft. I had it before[,] and [it is] no fun.'' P6 responded, ``I [do not] want the information compromised or sold to companies to stalk me.''

Nonetheless, in terms of understanding whether contact tracing is essential, the word `yes' arises as a significant term. Some lower-order terms include `spread' and `stop.' These terms collectively reveal an unexpected determination to leverage contact tracing as a way of \emph{curbing virus spread}. When asked if they believed in the importance of contact tracing, P7 replied, ``Absolutely - it is the only way to stop the spread.'' P8 responded, ``Yes if it's done right and safe to use for the users. It will immensely stop the disease.'' In the case of a HIV spread, P9 expressed, ``No, because I don't see much of it in my community.''

Our work aims to unpack the strong user resistance on digital contact tracing and discover better means of facilitating critical information sharing. These preliminary steps  hint toward promising insights into how information sharing in the contact tracing process can be reshaped to balance user concerns.
\begin{table*}[t]
\caption{Model summary for user willingness to share critical contact tracing information. }
\label{tab:modelsummary}
\centering
\resizebox{.87\linewidth}{!}{
\begin{tabular}{|c|c|p{50pt}|c|c|c|c|c|c|c|c|}\hline
\textbf{Model} & \textbf{DV} & \textbf{modality} & \textbf{demographic} & ${AIC}$ & ${R}^2$ & adj. ${R}^2$ & Std.${e}$ & ${df}$ & ${F}$ & ${Sig.}$ \\\hline

${M}^{identity}$ & {identity} & smartphone,\par PH-HR & {parenthood,income} & -293.45 & 0.819 & 0.816 & 0.501 & (4, 211) & 240.1 & 0`***' \\ \hline

${M}^{contact}$ & {contact details} & smartphone,\par PH-HR & {education,income,age} & -393.73 & 0.899 & 0.896 & 0.396 & (5, 210) & 374.4 & 0`***' \\ \hline

${M}^{location}$ & location & smartphone,\par PH-HR & parenthood, \par income,trust  & -278.88 & 0.857 & 0.853 & 0.461 & (5, 178) & 214.1 & 0`***' \\ \hline
 
${M}^{partner}$ & sex partner's details & smartphone,\par PH-HR & {education,income} & -51.98 & 0.903 & 0.889 & 0.413 & (4, 27) & 63.13 & 0`***' \\ \hline

\end{tabular} }     
\end{table*}

\section{Regression Analysis}
\label{sec:reg_analysis}
We carried out multiple linear regression to investigate the relationship between users' willingness to share private information for contact tracing (DV: dependent variables) and the preferred modality (IV: independent variables). We also utilized \emph{trust in public health} and a range of previously identified user demographics variables such as \emph{income} and \emph{education} as potential predictors.

\begin{figure}[h]
\centering
\includegraphics[width=\linewidth]{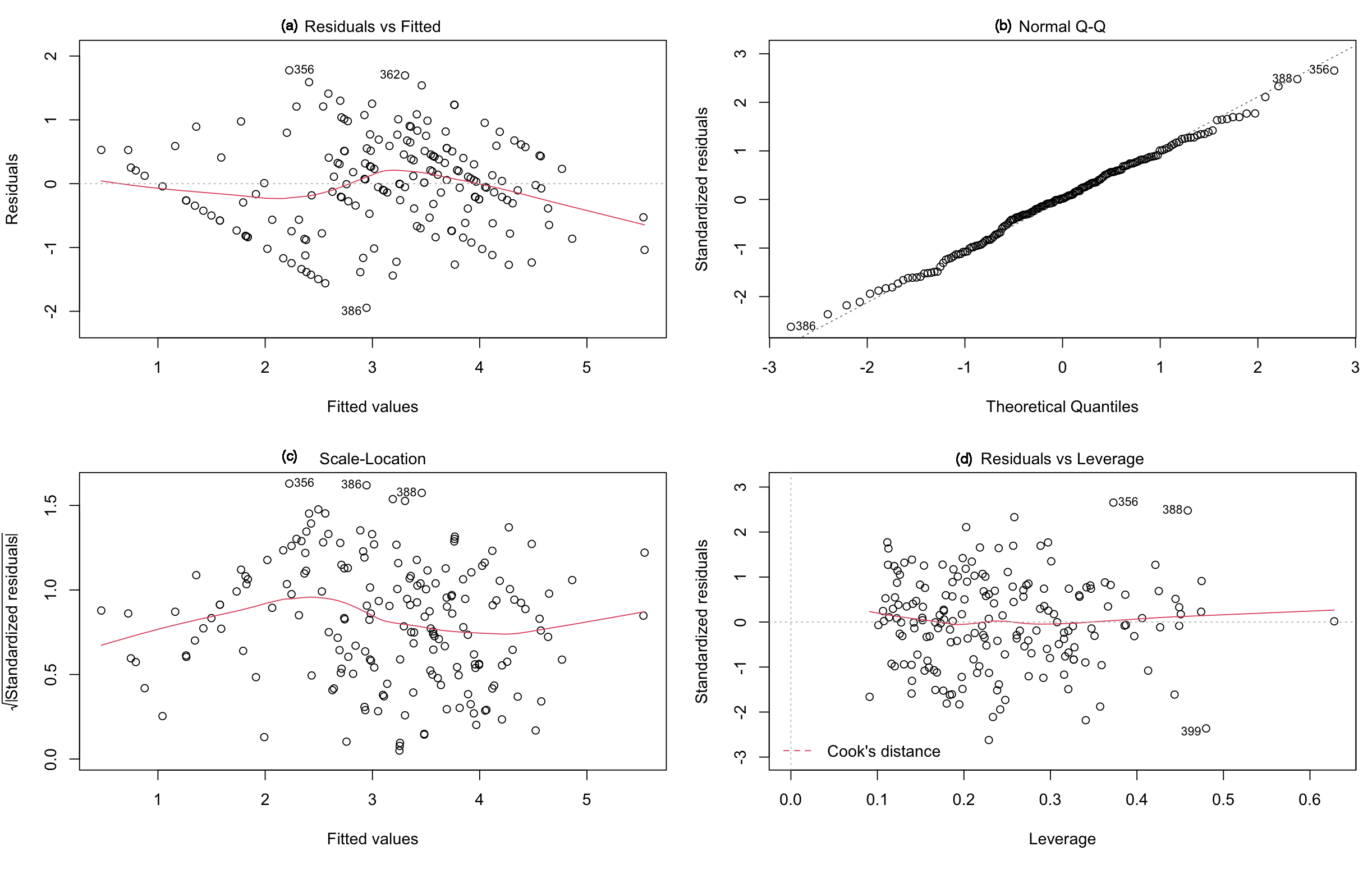}
\caption{Regression diagnostic.}
\label{fig:multireg_diagnostic}
\end{figure}

\begin{figure}[h]
\centering
\includegraphics[width=.7\linewidth]{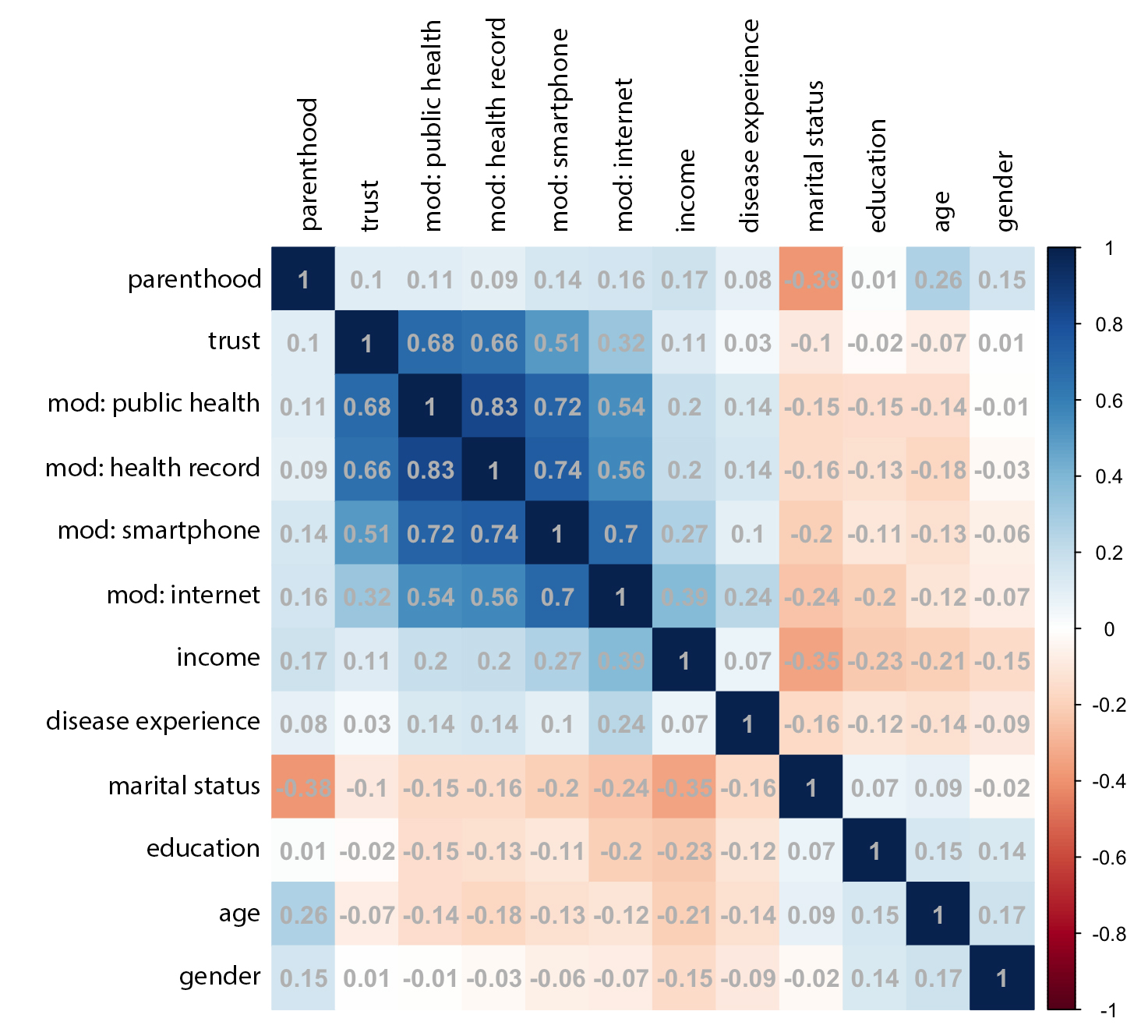}
\caption{Correlations between variables.}
\label{fig:multireg_corr}
\end{figure}

Our regression diagnostics plots in Figure \ref{fig:multireg_diagnostic} confirmed assumptions such as normality of errors, homoscedasticity, and multicollinearity were met for multiple regression to be reliable. The horizontal line (a) indicates a close to linear relationship without distinct patterns, the residual points follow the dashed line in (b), the horizontal line (c) indicates homoscedasticity, and (d) identifies few data points with high leverage. However, the standardized residuals of these observations are no more than 3 in absolute value as suggested by \cite{james2013introduction}.

The next step was to ensure no one independent variable is strongly correlated (r=0.8) with another. As common practice, highly correlated variables must be removed. Figure \ref{fig:multireg_corr} shows the correlations between demographic variables are no more than r=0.3. However, there is a high correlation between communicating with public health officers and retrieving health records for different modalities (r=0.83). 

We examined the phrasing of the questions (Questions 8.4 and 8.5 in Appendix \ref{appx:userstudy}), as well as the qualitative responses in order to understand this high correlation. There is a possibility that participants perceive the mode of collecting information ``personally from a public health official'' as bearing similarity with ``allowing such information to be collected from your doctor's medical health record.'' However, we are unable to unpack the meaning of this. Moving forward, we have decided to account for this high correlation by using the data for ``health records'' modality and to discuss the combined item as ``public health-health record'' (PH-HR).

We carried out stepwise regression to procedurally determine the best combination of independent variables (IV). Table \ref{tab:modelsummary} lists the model summary for each of our dependent variables (DV). We iteratively performed a backward selection to achieve a simplified model with the lowest Akaike Information Criteria (AIC), without impacting model performance; the lowest value quantifies the amount of information loss due to model simplification.

Overall, the results suggest that more than 80\% of the variance in the data can be explained by all models (${R}^2$ for ${M}^{location}$: 0.857, ${M}^{partner}$: 0.903, ${M}^{identity}$: 0.819, ${M}^{contact}$: 0.899). We summarize the coefficient estimates for all four models in Tables \ref{tab:coefficientLoc} to \ref{tab:coefficientIdn}. \\

\noindent\textbf{Choice of Modality} 
In these models, we found the use of smartphone and PH-HR consistent as significant predictors. Specifically in predicting user willingness to share location data (${M}^{location}$), the positive coefficient indicates an increase by 0.533 (\emph{$p<.001$}) given a one-unit shift in the use of smartphone with other variables held constant (see Table \ref{tab:coefficientLoc}). This result allows us to assess the effect of each modality isolated from other variables, confirming the use of smartphone to gather location information to be higher than other modalities.

In contrast, we found participants who responded to the HIV disease scenario to be more willing to share their sexual partner information through the use of PH-HR. As shown in Table \ref{tab:coefficientSxp}, the mean in willingness to provide sexual partner information (${M}^{partner}$) increases by 0.559 given a one-unit shift in the independent variable with the smartphone modality (and other variables) held constant. Our findings in Table \ref{tab:coefficientCon} indicate user willingness in utilizing PH-HR to extract contact information (${M}^{contact}$, $\beta$=0.510, $p<.001$). This result supports the feasibility in extracting characteristics of household contacts from the electronic medical record, as suggested by Metlay \etals \cite{metlay2021household}.

Finally, in Table \ref{tab:coefficientIdn}, our results seem to suggest that participants might be less particular of the modality choice ($\beta$ smartphone=0.449, $p<.001$, health record=0.489, $p<.001$) for providing information exposing their identity (${M}^{identity}$).\\

\noindent\textbf{Demographic Details.} 
In all models, users' \emph{income} is one demographic data that commonly yields significance. The positive coefficient indicates that as income increases, the mean of the sharing location, contact and personally identifiable information for contact tracing tend to increase. 

\emph{Education} is also a significant demographic predictor for sharing contact and sexual partner information. It should be noted that education level is an ordinal independent variable (i.e., 1: Skilled, 2: Degree, 3: Higher Degree). Education is observed to have a negative relationship in these models, suggesting that users with lower educational attainment are less willing to provide contact tracing information.

\begin{table}[t]
\captionof{table}{Coefficient summary to predict user willingness to share location information, ${M}^{location}$.}
\label{tab:coefficientLoc}
\resizebox{.78\linewidth}{!}{
\begin{tabular}{|l|c|c|c|c|l|}\hline
\textbf{Coefficient} & $\beta$ & ${est.}$ & Std.${e}$ & ${t}$ & ${Sig.}$ \\\hline
(Intercept)       &&-0.09 &0.160  &-0.563  &0.573    \\
smartphone	      &0.533  &0.483    &0.038  &12.686  &< 2e-16 ***\\
PH-HR			  &0.369  &0.338    &0.043  &7.739   &7.23e-13 ***\\
parenthood        &-0.048 &-0.053   &0.031  &-1.677  &0.095 .  \\
income            &0.082  &0.087    &0.031  &2.802   &0.005 ** \\
trust	          &0.100  &0.129    &0.050  &2.562   &0.011 * \\\hline
\end{tabular}}
\end{table}

\begin{table}[t]
\caption{Coefficient summary to predict user willingness to share sexual partner information, ${M}^{partner}$.}
\label{tab:coefficientSxp}
\resizebox{.78\linewidth}{!}{
\begin{tabular}{|l|c|c|c|c|l|}\hline
\textbf{Coefficient} & $\beta$ & ${est.}$ & Std.${e}$ & ${t}$ & ${Sig.}$ \\\hline
(Intercept)        &&1.083    &0.370   &2.927  &0.006 **  \\
smartphone    	   &0.419     &0.380    &0.100   &3.805  &0.000 *** \\
PH-HR  			   &0.559     &0.479    &0.091   &5.222  &1.68e-05 *** \\
education          &-0.184    &-0.218   &0.083   &-2.608 &0.014 *   \\
income             &-0.163    &-0.169   &0.075   &-2.245 &0.033 * \\\hline
\end{tabular}}     
\end{table}

\begin{table}[t]
\caption{Coefficient summary to predict user willingness to share contact information, ${M}^{contact}$.}
\label{tab:coefficientCon}
\resizebox{.78\linewidth}{!}{
\begin{tabular}{|l|c|c|c|c|l|}\hline
\textbf{Coefficient} & $\beta$ & ${est.}$ & Std.${e}$ & ${t}$ & ${Sig.}$ \\\hline
(Intercept)        &&0.040    &0.140   &0.286  &0.775  \\   
smartphone	       &0.485	  &0.445    &0.030   &14.576 &< 2e-16 *** \\ 
PH-HR			   &0.510	  &0.468    &0.030   &15.438 &< 2e-16 *** \\
education          &-0.063	  &-0.077   &0.027   &-2.817 &0.005 **  \\
income             &0.052	  &0.055    &0.025   &2.216  &0.027 *   \\
age                &0.045	  &0.042    &0.021   &1.994  &0.047 *  \\\hline
\end{tabular}}
\end{table}

\begin{table}[t]
\caption{Coefficient summary to predict user willingness to share personal identity, ${M}^{identity}$.}
\label{tab:coefficientIdn}
\centering
\resizebox{.78\linewidth}{!}{
\begin{tabular}{|l|c|c|c|c|l|}\hline
\textbf{Coefficient} & $\beta$ & ${est.}$ & Std.${e}$ & ${t}$ & ${Sig.}$ \\\hline
(Intercept)       &&0.005   &0.120   &0.043   &0.966    \\
smartphone   	  &0.449    &0.390   &0.038   &10.084  &<2e-16 ***\\
PH-HR			  &0.489    &0.426   &0.038   &11.207  &<2e-16 ***\\
parenthood        &0.042    &0.046   &0.032   &1.441   &0.1511    \\
income            &0.075    &0.076   &0.031   &2.474   &0.014 * \\\hline
\end{tabular}} 
\end{table}

Separately, \emph{parenthood} reveals inverse relationships between sharing location and personal particulars. Non-parent users are less willing to provide location data, while parents are more willing to provide personal data for contact tracing.

Finally, it is worth noting that \emph{trust in public health} plays a significant role in users providing location information, but not other kinds of information. The mean willingness to provide location data increases by 0.1, given a one-unit shift in trust as a predictor (with all else held constant).
We reviewed codes for open-ended answers in response to the question, ``What are your concerns with sharing contact tracing information for the disease?'' Respondents unwilling to share location information (1 on the Likert scale) seemed to be most concerned with privacy issues and concerns arising from cybercrimes or abuse of power. For example, P10 said, ``It might be misused and the data might reach third parties and endanger my privacy.'' P11 perceived contact tracing as a ``trojan horse for access[ing] private information of location, activities, and constitutional rights being infringed.''

\subsection{Summary of Findings}
In this section, we ran a quantitative analysis to explore the predictors affecting users' willingness to share their private information for contact tracing. Our results showed that leveraging internet browsing histories as a data collection modality is a redundant predictor to collect any information (see Table \ref{tab:modelsummary}). The evidence indicates strong support for using smartphones. The role of trust in public health organizations is particularly critical when users are asked to share location data through their smartphones.

Furthermore, this analysis suggests the importance of a modality \emph{PH-HR}, that may be a configuration of public health officials extracting relevant data from medical health records. The preference for sharing more private information such as contact data (for all disease types) and sexual partner information (for HIV) is higher through this modality. 

Regardless of data collection modality, some previously documented demographic, non-modifiable factors were found to significantly influence user willingness to provide these data. They are primarily \emph{income} and \emph{education} \cite{jonker2020covid}. \emph{Parenthood} emerged as an intriguing possible new demographic predictor. 
\section{Discussion}
\label{sec:discussion}
Our study's objectives were to better understand how individual factors and modes of data collection affect user acceptance in digital contact tracing. Here we discuss the key takeaways of our findings.

\subsection{How Mode Influences User Willingness to Share Contact Tracing Information}
\label{sec:discussion1}
Since the COVID-19 outbreak, much focus on mobilizing contact tracing efforts has fundamentally relied on communicating with public health officials and utilizing smartphone sensing modalities for partial automation of contact tracing. Our study sought to examine these and alternative methods of data collection. 

\subsubsection{\revise{Influence of Disease Type on Contact Tracing Preferences}}
\revise{While our study observed six different diseases, each depicting a unique feature spread, our statistical findings did not demonstrate strong evidence of disease groups influencing users' willingness to share different types of critical contact tracing. Accordingly, we had no basis to pursue our regression analysis by disease type. It is important to note that our study did not investigate the threat perception of each disease, which may have exhibited differentiation \cite{perez2020questionnaire,sadique2007precautionary}. Further, we cannot determine if the lack of differentiation by disease resulted from studying users' responses to the mode of contact tracing efforts and not to each unique disease. Future work may require a systematic examination of threat perception for different disease states and whether it is a moderating factor in users' responses to the mode of digital contact tracing.}

\subsubsection{User Resistance towards Internet Modality}
Despite some success in using automated data collection of search results, our findings suggest that acceptability will be low. In a study related to COVID-19, Li \etals highlights internet data as a reliable source that can accurately predict the COVID-19 outbreak one to two weeks earlier before the first laboratory confirmation was announced. Word searches such as `coronavirus' and `pneumonia' peaked on Google Trend \cite{li2020retrospective} five days before the COVID-19 was announced as a global health emergency by WHO \cite{whoNews}. Similarly, Shin \etals found internet data to monitor MERS-CoV (Middle East respiratory syndrome) three days before laboratory confirmations \cite{shin2016high}. Accordingly, prior work has proposed internet surveillance data to complement tracking infection trends in the COVID-19 pandemic \cite{lu2021internet}. In this work, we examined the idea of collecting contact tracing information through internet browsing histories as a possible modality. As shown in Figure \ref{fig:infoModalitiesHiv}, utilizing search data to gather any contact tracing information is not likely to  reach beyond a 20\% adoption, thus, insufficiently sized for mass adoption. This result is not surprising, given high concerns over behavioral advertising through surreptitious tracking and targeting of users. Such utilization of internet data is precisely expressed as a concern. P12 explained, ``My concerns would be using the internet as a source of information. It would give access to information that does not pertain to health.''

\subsubsection{Potential of Medical Health Records in Coordination with Public Health Personnel as a Modality}
Our findings support the use of medical health records for collecting contact tracing information. Specifically, our exploratory analysis found approximately 75\% users willing to share contact data and close to 90\% users willing to provide personal information through medical health records. Indeed, this approach was reported by Metlay \etals to be effective in assisting with COVID-19 contact tracing. The authors successfully identified immediate contacts of a patient through using electronic medical health records, leading to crucial information for estimating household transmission risk \cite{metlay2021household}. However, medical health records are generally a poor source of current location information. There may be outdated home addresses or miscounting of immediate family contacts. 
However, since our findings suggest utilizing medical health records enjoys high user acceptance as a data source for gathering contact and personal information, it may be worth discussing further how to employ this mode of digital contact tracing. 

\subsection{Users' Concerns within Each Mode}
Our regression analysis revealed user's perception of different modalities and the activity of information exchange.

\subsubsection{Smartphone Location-sensing Needs Trust}
Our participants were more willing to share location data when the mode of digital contact tracing was smartphones. However, this finding does not adhere with the low application uptake for digital contact tracing mobile apps deployed in the U.S. during the COVID-19 pandemic\cite{timesCT}. These systems relied on alert notification systems (for example, a joint initiative by Apple and Google) that solely track exposure through utilizing Bluetooth proximity \cite{aphl2020, nashp2020}. 

Our study cannot precisely explain the disjunct between the stated preference to use a smartphone in sharing location data and the current poor acceptance of digital contact tracing. The open-ended answers suggests that this push back is concerned with restrictions on individual liberties and with personal safety. P13 asserted, ``sharing my personal whereabouts with [the] government [will infringe on] my freedoms.'' P14 said, ``[Tracking location] will lead to restrictions in people's ability to move about and go places once controls are implemented.''  Finally, P15 explained, ``I am concerned about the [metadata] that can be gleaned from contact tracing apps then later used against me. What if I am circumstantially somewhere where a crime is committed, and my cell phone was the only one in the area [\emph{sic}] I just [do not] want government having my [metadata] even though they can probably get it anyway.'' 

At the same time, our findings allude to the significance of trust in public health officials in moderating these intuitive concerns \cite{phelan2016s} of privacy on digital contact tracing. Specifically, users with higher trust in public health officials may be more willing to supply their location data\cite{hassandoust2021individuals,nurgalieva2021attitudes}. Additionally, we learned from our qualitative findings that more participants had communicated confidence in public health agencies to lead the contact tracing effort. P16 expressed, ``I do not have any worry about sharing my contact tracing information because I know the public health officials have my best interest at heart.''

\subsubsection{Addressing Limitations of Direct Communication with a Public Health Official}
In most scenarios, the users preferred to communicate directly with a public health official. However, in the HIV scenario, users were more cautious towards sharing sensitive sexual partner and contact information. This difference is not statistically significant but suggests different sensitivities might need to be addressed. Additionally, direct contact by public health officials in the US is complicated by a majority of Americans screening unknown calls and voicemails~\cite{trust2020pew}. Manual contact tracing is also limited by the difficulty of scaling up. 

In the U.S. today, the mechanism of authorizing information sharing with primary care physicians is recommended for COVID-19 vaccine administration to maintain patient safety~\cite{aafp2020}. We suggest the same kind of information sharing could be modified to allow for digital contact tracing. Upon consent, a user could preemptively authorize public health officials to access their medical records by providing information of the primary care physician. 

\subsection{Influence of Demographics on the Type of Information in Each Mode}
Strategizing for high community engagement is a critical component for whole nation contact tracing to be successful. CDC and WHO have emphasized the need to improve marketing campaigns to targeted user groups \cite{cdc2020,world2020ethical}. In this discussion point, we use our findings to shed light on the potential hard-to-reach groups.

\subsubsection{Socioeconomic status (SES)}
Our regression results yield income as a significant demographic factor in all models. It suggests that lower income users' can be less willing to participate. This finding corresponds with prior work that users with lower-income and less formal education tend to be less comfortable sharing information \cite{mcclain2020challenges}. This effect seemed to be particularly strong in sharing sexual partner information under an HIV outbreak. 

Recent work about contact tracing in Australia~\cite{Lockey2021} suggests the age of the user may further moderate the effect of socioeconomic status, where older users with low socioeconomic status are willing to participate in digital contact tracing tools. Separately, a cross-section study in France found a positive correlation between acceptance of digital contact tracing and one's health literacy level in understanding the disease \cite{touzani2021mhealth}. While our study did not study the effects of health literacy, much research has revealed the significance of SES inequalities, including lower-income and less formal education, to negatively impact one's understanding of basic health information and its appropriate services \cite{rikard2016examining}. Roughly one-half of American adults were reported to exhibit low health literacy in the U.S. \cite{mackert2016jmir}, but the same study promisingly indicates a positive association with trust in public health. More work is required to understand if the presence of mechanisms of co-production of trust, as suggested by \cite{megnin2020facilitators} may be particularly useful among lower SES users. One key takeaway is that public health messaging within lower SES communities may focus on basic disease information and privacy assurance when using digital contact tracing tools.

\subsubsection{Domestic Characteristics and Parenthood}
Previous work has identified demographic characteristics that influence contact tracing adoption \cite{Lockey2021,jonker2020covid}. Our findings add to this body of work by suggesting parenthood as a new facilitator.

A household is considered a significant venue for transmission in a pandemic situation, putting everyone under the same roof at high risk \cite{lewis2020household,metlay2021household}. With contact tracing bringing higher health security for all, our findings suggest that a user with family members to consider could display a more significant commitment to following safety measures, including embracing contact tracing mechanisms. This finding may also further contextualize \cite{megnin2020facilitators} findings regarding perceived responsibility and perceived benefit as facilitators of acceptance of contact tracing efforts. 

\subsubsection{Messaging Opportunities for Promoting Digital Contact Tracing}
Our results identified demographic predictors of willingness to share information such as parenthood, education, and income.  These results could also be used to inform the messaging for contact tracing initiatives.
For example, lower education is a predictor of users' unwillingness to share contact, identity, and sexual partner information. In the case of an outbreak of diseases where such information is needed (e.g., STIs), contact tracing campaigns may need to enhance outreach among populations with lower education or lower literacy levels. On the other hand, parenthood showed promise in terms of predicting willingness to share information. In the case of COVID-19, while young children tend to show milder symptoms~\cite{Lee2020}, starting digital contact tracing through K-12 schools could be a good way to start a larger-scale initiative on a community level. Parents would be indirectly targeted by the initiative, hopefully ensuring a more extensive user base in the long run. Due to the variances in willingness to share information amongst different population groups, targeted messaging could be a promising method for promoting the uptake of digital contact tracing.

\section{Limitations and Future Work}
\label{sec:futurework}
\revise{As depicted in Appendix \ref{appx:exploratory}, although we do not find statistical differences in the willingness to participate in contact tracing by different disease groups, we visually observe certain tendencies in some disease groups compared to others. Our approach to combining different groups into a single analysis may affect the differences in each disease group. Future work should identify the nuances of each disease scenario and understand how these nuances influence the user's willingness to participate in contact tracing.}

Our work was done through Amazon Mechanical Turk, with a specific focus on the U.S. As such, the results might not be generalizable to other countries with significant cultural differences. The survey is limited by the nature of self-reported data, where responses are based on a hypothetical scenario. Future work should systematically assess users' acceptance of a widespread digital contact tracing solution. Some critical factors include users' demographics, perception, and health literacy.

Another limitation is our respondents' overall high level of trust due to the platform used to collect data. By definition, crowd workers are more willing to provide information online. Future work may be required to replicate this survey using paper-and-pen methods. Finally, our study was limited to sharing critical but sensitive information as per CDC guidelines. The transformation of such information from manual practice to digital forms could adopt privacy-preserving operations. With the evolving role of digital technologies, future work should alternatively investigate users' willingness to use different modes for anonymous tracing. A deeper qualitative evaluation is required to better understand users' attitudes towards these modalities.
\section{Conclusion}
\label{sec:conclusion}
We conducted a survey study where more than 200 respondents from the U.S. participated in one of six infectious disease scenarios (i.e., HIV, nCov, Zika, Ebola, MRSA, and Hep), listed as top global health threats by the WHO. The findings are similar across these diseases, suggesting strong generalizability of results beyond the current COVID-19 pandemic. We have presented our first steps in identifying the facilitators and barriers of contact tracing in a U.S. setting. Our study is set apart from prior work by examining user perception of contact tracing across six different disease types -- without the procedure tied to a particular application, but the mode of information exchange itself. First, the study showed how user willingness to provide contact tracing information could be improved by incorporating various modalities, specifically medical health records and smartphones. Second, trust in public health agencies is moderately high, suggesting a more coordinated effort of digital contact tracing overseen by public health institutions rather than government and technology companies in the U.S. Third, and finally, the study shed more light on the types of individual demographic factors that need targeted strategies to improve contact tracing uptake.

\begin{acks}
This work was supported by grants of the SGP Healthcare Fund, PIE-SGP-HC-2019-01 (Thrust 3-2). Any opinions, findings, conclusions, or recommendations expressed in this paper are those of the authors and do not necessarily reflect the views of the funding agency.
\end{acks}

\bibliographystyle{ACM-Reference-Format}
\bibliography{ACM-Reference-Format}

\appendix
\section{Scenario-based survey: Full Questionnaire}
\label{appx:userstudy}

Q1. \textbf{Novel coronavirus (nCov)} can be transmitted directly from a person to another.\\
$\bigcirc{}$  True\\
$\bigcirc{}$  False\\

Q2. Based on the information needs highlighted in yellow above, how willing would you be in allowing such information to be collected \textbf{automatically from your smartphone}?

\scalebox{.7}
{\begin{tabular}{llllll}
& Strongly  &  &  &  &  Strongly  \\
&  Disagree & Disagree & Undecided & Agree &  Agree \\
Your Identity & \multicolumn{1}{c}{$\bigcirc{}$}  & \multicolumn{1}{c}{$\bigcirc{}$}  & \multicolumn{1}{c}{$\bigcirc{}$}  & \multicolumn{1}{c}{$\bigcirc{}$} & \multicolumn{1}{c}{$\bigcirc{}$}  \\
Contact Data & \multicolumn{1}{c}{$\bigcirc{}$}  & \multicolumn{1}{c}{$\bigcirc{}$}  & \multicolumn{1}{c}{$\bigcirc{}$}  & \multicolumn{1}{c}{$\bigcirc{}$} & \multicolumn{1}{c}{$\bigcirc{}$} \\
Location of places
& \multicolumn{1}{c}{$\bigcirc{}$}  & \multicolumn{1}{c}{$\bigcirc{}$}  & \multicolumn{1}{c}{$\bigcirc{}$}  & \multicolumn{1}{c}{$\bigcirc{}$} & \multicolumn{1}{c}{$\bigcirc{}$} \\
you have visited \\
\end{tabular}}
\newline

Q3. Based on the information needs highlighted in yellow above, how willing would you be in allowing such information to be collected \textbf{automatically from your doctor's medical health record}?

\scalebox{.7}
{\begin{tabular}{llllll}
& Strongly  &  &  &  &  Strongly  \\
&  Disagree & Disagree & Undecided & Agree &  Agree \\
Your Identity & \multicolumn{1}{c}{$\bigcirc{}$}  & \multicolumn{1}{c}{$\bigcirc{}$}  & \multicolumn{1}{c}{$\bigcirc{}$}  & \multicolumn{1}{c}{$\bigcirc{}$} & \multicolumn{1}{c}{$\bigcirc{}$}  \\
Contact Data & \multicolumn{1}{c}{$\bigcirc{}$}  & \multicolumn{1}{c}{$\bigcirc{}$}  & \multicolumn{1}{c}{$\bigcirc{}$}  & \multicolumn{1}{c}{$\bigcirc{}$} & \multicolumn{1}{c}{$\bigcirc{}$} \\
Location of places
& \multicolumn{1}{c}{$\bigcirc{}$}  & \multicolumn{1}{c}{$\bigcirc{}$}  & \multicolumn{1}{c}{$\bigcirc{}$}  & \multicolumn{1}{c}{$\bigcirc{}$} & \multicolumn{1}{c}{$\bigcirc{}$} \\
you have visited \\
\end{tabular}}
\newline

Q4. Based on the information needs highlighted in yellow above, how willing would you be in allowing such information to be collected \textbf{automatically from your Internet browsing activities}?

\scalebox{.7}
{\begin{tabular}{llllll}
& Strongly  &  &  &  &  Strongly  \\
&  Disagree & Disagree & Undecided & Agree &  Agree \\
Your Identity & \multicolumn{1}{c}{$\bigcirc{}$}  & \multicolumn{1}{c}{$\bigcirc{}$}  & \multicolumn{1}{c}{$\bigcirc{}$}  & \multicolumn{1}{c}{$\bigcirc{}$} & \multicolumn{1}{c}{$\bigcirc{}$}  \\
Contact Data & \multicolumn{1}{c}{$\bigcirc{}$}  & \multicolumn{1}{c}{$\bigcirc{}$}  & \multicolumn{1}{c}{$\bigcirc{}$}  & \multicolumn{1}{c}{$\bigcirc{}$} & \multicolumn{1}{c}{$\bigcirc{}$} \\
Location of places
& \multicolumn{1}{c}{$\bigcirc{}$}  & \multicolumn{1}{c}{$\bigcirc{}$}  & \multicolumn{1}{c}{$\bigcirc{}$}  & \multicolumn{1}{c}{$\bigcirc{}$} & \multicolumn{1}{c}{$\bigcirc{}$} \\
you have visited \\
\end{tabular}}
\newline

Q5. Based on the information needs highlighted in yellow above, how willing would you be in allowing such information to be collected \textbf{personally from a public health official}?

\scalebox{.7}
{\begin{tabular}{llllll}
& Strongly  &  &  &  &  Strongly  \\
&  Disagree & Disagree & Undecided & Agree &  Agree \\
Your Identity & \multicolumn{1}{c}{$\bigcirc{}$}  & \multicolumn{1}{c}{$\bigcirc{}$}  & \multicolumn{1}{c}{$\bigcirc{}$}  & \multicolumn{1}{c}{$\bigcirc{}$} & \multicolumn{1}{c}{$\bigcirc{}$}  \\
Contact Data & \multicolumn{1}{c}{$\bigcirc{}$}  & \multicolumn{1}{c}{$\bigcirc{}$}  & \multicolumn{1}{c}{$\bigcirc{}$}  & \multicolumn{1}{c}{$\bigcirc{}$} & \multicolumn{1}{c}{$\bigcirc{}$} \\
Location of places
& \multicolumn{1}{c}{$\bigcirc{}$}  & \multicolumn{1}{c}{$\bigcirc{}$}  & \multicolumn{1}{c}{$\bigcirc{}$}  & \multicolumn{1}{c}{$\bigcirc{}$} & \multicolumn{1}{c}{$\bigcirc{}$} \\
you have visited \\
\end{tabular}}
\newline

Q6. Please state your most preferred way of providing the following information:

\begin{tabular}{ll}
 & \multicolumn{1}{c}{Most Preferred Access Method} \\
Your Identity & \multicolumn{1}{c}{\fbox{Select $\nabla$ }} \\
Contact Data & \multicolumn{1}{c}{\fbox{Select $\nabla$ }} \\
Location Visited & \multicolumn{1}{c}{\fbox{Select $\nabla$ }}\\
\end{tabular}%
\newline

Q7. What is your experience with \textbf{novel coronavirus (nCov)}? (You may select more than one option) \\
$\square{}$  Personally\\
$\square{}$  Friends/family(close circle)\\
$\square{}$  None\\

Q8. Using the following scale, please indicate how much you agree or disagree with the following statements:

\scalebox{.75}{
\begin{tabular}{p{100pt}lllll}
& Strongly  &  &  &  &  Strongly  \\
&  Disagree & Disagree & Undecided & Agree &  Agree \\
I believe that the public\par health officials would\par act in my best interest. & \multicolumn{1}{c}{$\bigcirc{}$}  & \multicolumn{1}{c}{$\bigcirc{}$}  & \multicolumn{1}{c}{$\bigcirc{}$}  & \multicolumn{1}{c}{$\bigcirc{}$} & \multicolumn{1}{c}{$\bigcirc{}$}  \\
If I required help, the\par public health officials\par would do their best to\par help me. & \multicolumn{1}{c}{$\bigcirc{}$}  & \multicolumn{1}{c}{$\bigcirc{}$}  & \multicolumn{1}{c}{$\bigcirc{}$}  & \multicolumn{1}{c}{$\bigcirc{}$} & \multicolumn{1}{c}{$\bigcirc{}$} \\
The public health officials\par  are interested in my well\par -being, not just their own. & \multicolumn{1}{c}{$\bigcirc{}$}  & \multicolumn{1}{c}{$\bigcirc{}$}  & \multicolumn{1}{c}{$\bigcirc{}$}  & \multicolumn{1}{c}{$\bigcirc{}$} & \multicolumn{1}{c}{$\bigcirc{}$} \\
The public health officials\par are truthful in their\par dealings with me. & \multicolumn{1}{c}{$\bigcirc{}$}  & \multicolumn{1}{c}{$\bigcirc{}$}  & \multicolumn{1}{c}{$\bigcirc{}$}  & \multicolumn{1}{c}{$\bigcirc{}$} & \multicolumn{1}{c}{$\bigcirc{}$}  \\
I would characterise the\par public health officials as\par honest. & \multicolumn{1}{c}{$\bigcirc{}$}  & \multicolumn{1}{c}{$\bigcirc{}$}  & \multicolumn{1}{c}{$\bigcirc{}$}  & \multicolumn{1}{c}{$\bigcirc{}$} & \multicolumn{1}{c}{$\bigcirc{}$} \\
The public health officials\par would keep their\par commitments. & \multicolumn{1}{c}{$\bigcirc{}$}  & \multicolumn{1}{c}{$\bigcirc{}$}  & \multicolumn{1}{c}{$\bigcirc{}$}  & \multicolumn{1}{c}{$\bigcirc{}$} & \multicolumn{1}{c}{$\bigcirc{}$} \\
The public health officials\par are sincere and genuine. & \multicolumn{1}{c}{$\bigcirc{}$}  & \multicolumn{1}{c}{$\bigcirc{}$}  & \multicolumn{1}{c}{$\bigcirc{}$}  & \multicolumn{1}{c}{$\bigcirc{}$} & \multicolumn{1}{c}{$\bigcirc{}$} \\
The public health officials\par are competent and effective\par in providing public health\par and safety advice. & \multicolumn{1}{c}{$\bigcirc{}$}  & \multicolumn{1}{c}{$\bigcirc{}$}  & \multicolumn{1}{c}{$\bigcirc{}$}  & \multicolumn{1}{c}{$\bigcirc{}$} & \multicolumn{1}{c}{$\bigcirc{}$}  \\
The public health officials\par perform their role of giving\par public health and safety\par  advice very well. & \multicolumn{1}{c}{$\bigcirc{}$}  & \multicolumn{1}{c}{$\bigcirc{}$}  & \multicolumn{1}{c}{$\bigcirc{}$}  & \multicolumn{1}{c}{$\bigcirc{}$} & \multicolumn{1}{c}{$\bigcirc{}$} \\
Overall, the public health\par officials are capable and\par proficient safety providers. & \multicolumn{1}{c}{$\bigcirc{}$}  & \multicolumn{1}{c}{$\bigcirc{}$}  & \multicolumn{1}{c}{$\bigcirc{}$}  & \multicolumn{1}{c}{$\bigcirc{}$} & \multicolumn{1}{c}{$\bigcirc{}$} \\
In general, the public\par health officials are very\par knowledgeable about public\par  health and safety. & \multicolumn{1}{c}{$\bigcirc{}$}  & \multicolumn{1}{c}{$\bigcirc{}$}  & \multicolumn{1}{c}{$\bigcirc{}$}  & \multicolumn{1}{c}{$\bigcirc{}$} & \multicolumn{1}{c}{$\bigcirc{}$}
\end{tabular}
}
\newline

Q9. How concerned are you with \textbf{novel coronavirus(nCov)}? \\
$\bigcirc{}$  Low\\
$\bigcirc{}$  Medium\\
$\bigcirc{}$  High\\
\newline

Q10. What are your concerns with sharing contact tracing information for \textbf{novel coronavirus (nCov)}?\\
\rule{8.5cm}{0.15mm}\\

Q11. What concerns you most about an outbreak of \textbf{novel coronavirus (nCov)}?\\
\rule{8.5cm}{0.15mm}\\

Q12. Do you believe contact tracing is important for \textbf{novel coronavirus (nCov)}? Why?\\
\rule{8.5cm}{0.15mm}

\newpage
\section{Codebook for Open-ended Responses}
\label{appx:codebook}

\begin{table}[h!]
\resizebox{\linewidth}{!}{\begin{tabular}{|p{90pt}|p{225pt}|} 
\hline
\textbf{Code}                              & \textbf{Code Definition}\\ 
\hline
Data Leakage/Theft                         & Respondent is fearful of data being leaked to unauthorised\par parties \\ 
\hline
Identity Theft                             & Respondent is concerned that their identity might be stolen\\ 
\hline
Misuse of Data                             & Respondent is fearful of data used for unauthorized reasons\\ 
\hline
Trust - Negative                           & Respondent is concerned about trust towards entity collecting\par the data\\ 
\hline
Trust - Positive                           & Respondent trusts entity collecting the data\\ 
\hline
Loss of Privacy                            & Respondent is concerned about loss of personal privacy\\ 
\hline
Data Confidentiality                       & Respondent is concerned about how the data is stored\par confidentially\\ 
\hline
No Concern                                 & Respondent demonstrates no concerns about sharing\par contact tracing information\\ 
\hline
Surveillance                               & Respondent is fearful of data used for surveillance purposes\\ 
\hline
Stigma                                     & Respondent is concerned about stigma of contracting the\par disease\\ 
\hline
Financial implications\par to self            & Respondent is fearful of loss of job, income or livelihood\\ 
\hline
Financial implications\par to society          & Respondent demonstrates concerns about widespread\par economic downturn due to disease \\ 
\hline
Fear of contracting\par disease                & Respondent demonstrates fear of contracting disease\\ 
\hline
Fear of death from\par disease                 & Respondent demonstrates fear of dying from the disease\\ 
\hline
Fear of family members\par contracting disease & Respondent demonstrates fear of family members\par contracting disease\\ 
\hline
Fear of spreading disease\par to social circle  & Respondent demonstrates fear of contracting and\par spreading the disease to people in social circle\\ 
\hline
No Concern                                 & Respondent demonstrates no fear of disease\\ 
\hline
Misinformation                             & Respondent demonstrates concerns about \par misinformation due to disease\\ 
\hline
Fear of Disease Control                    & Respondent demonstrates fear over how well\par the disease can be controlled \\ 
\hline
Supportive                                 & Respondent is supportive of contract tracing\\ 
\hline
Not Supportive                             & Respondent is not supportive of contract tracing\\ 
\hline
Unsure/Subjective                          & Respondent is unsure about effectiveness of\par contact tracing or would want more convincing reasons \\ 
\hline
Warns                                      & Respondent feels that contact tracing can help\par to warn about virus spread\\ 
\hline
Curb Virus Spread                          & Respondent feels that contact tracing can help to\par curb virus spread \\ 
\hline
Privacy Concerns                           & Respondent is concerned about sharing personal data\par for contact tracing\\ 
\hline
Public Health benefit                      & Respondent feels that contact tracing is beneficial\par for public health\\ 
\hline
Doesn't concern\par individual                & Respondent feels that contact tracing isn't\par important as the disease doesn't impact the\par individual or is not serious \\ 
\hline
Unlikely contraction                       & Respondent feels that it's unlikely to get infected. \\ 
\hline
Politics                                   & Respondent believes that the virus can be used for\par political reasons \\ 
\hline
Effectiveness/Infectivity                  & Respondent is skeptical about effectiveness/infectivity                                                               \\
\hline
\end{tabular}}
\end{table}
\newpage
\section{Pairwise Comparison Between Disease Groups}
\label{appx:exploratory}

\begin{figure}[h]
    \centering
    \includegraphics[width=.8\linewidth]{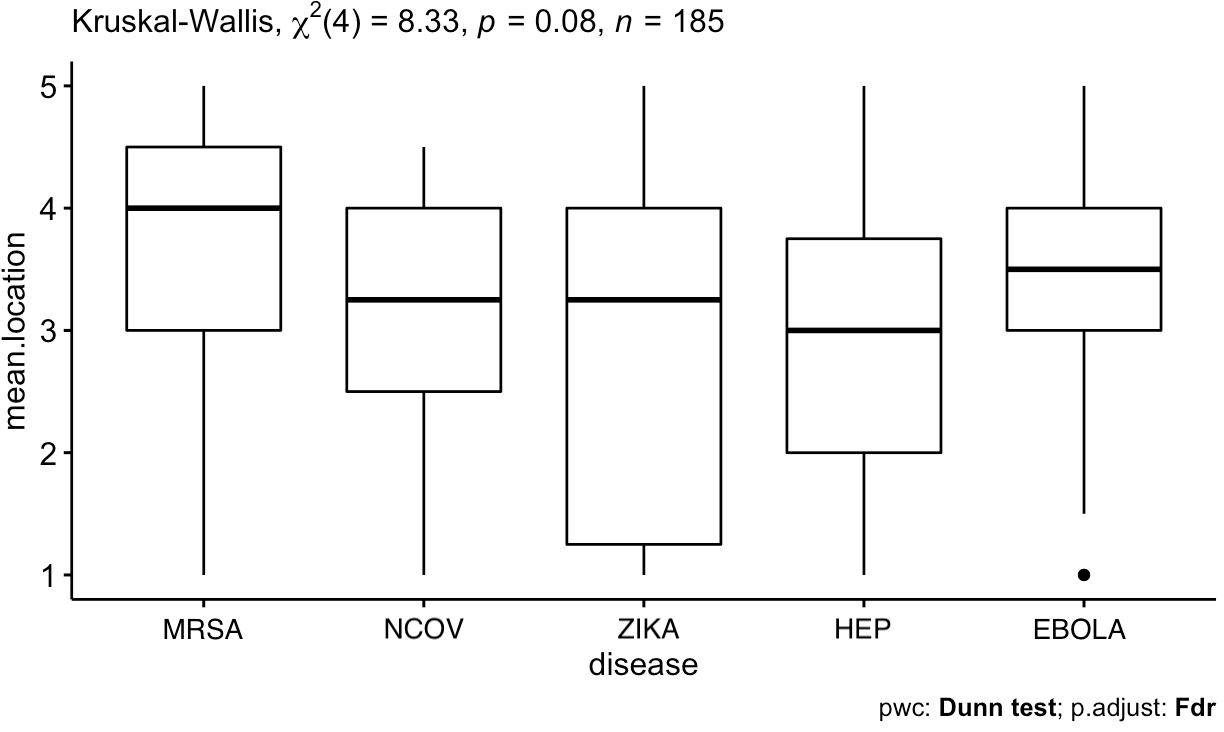}
    \caption{Sharing location data by disease}
    \label{tab:locationPlot}
\end{figure}

\begin{figure}[h]
    \centering
    \includegraphics[width=.85\linewidth]{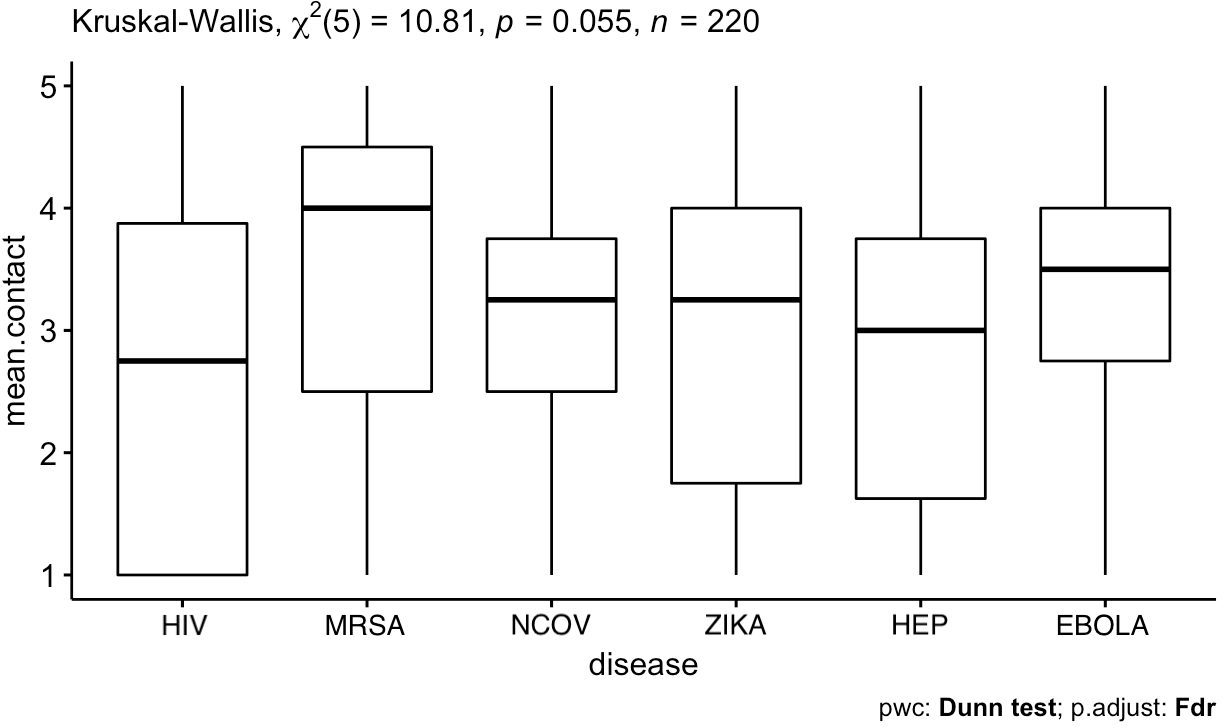}
    \caption{Sharing contact data by disease}
    \label{tab:contactPlot}
\end{figure}

\begin{figure}[h]
    \centering
    \includegraphics[width=.8\linewidth]{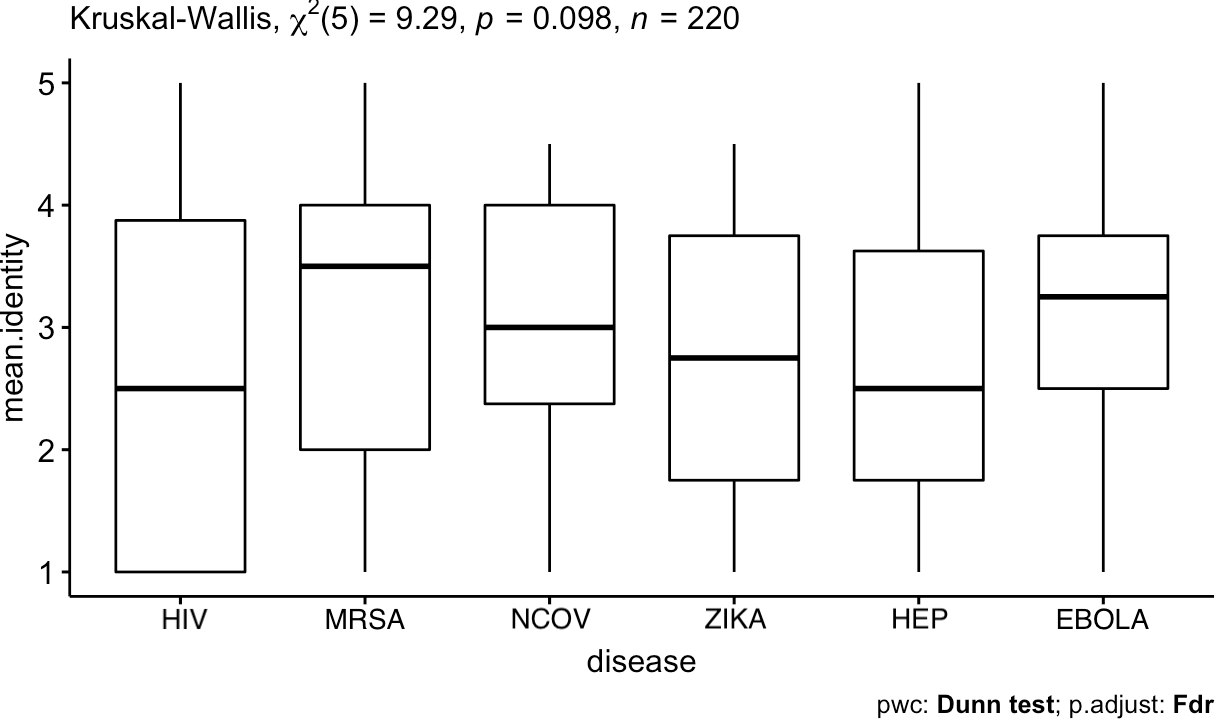}
    \caption{Sharing identity data by disease}
    \label{tab:identityPlot}
\end{figure}

\newpage
\begin{table}[h]
\caption{\revise{Dunn's test p-value summary for location}}
\label{tab:locationDunn}
\resizebox{\linewidth}{!}{
\begin{tabular}{|l|l|l|l|l|l|l|l|}\hline
\toprule
\textbf{Group 1} & \textbf{Group 2} & \textbf{n1} & \textbf{n2} & \textbf{statistic} & \textbf{p} & \textbf{p.adj} & \textbf{Sig} \\
\toprule

MRSA & nCov & 37 & 43 & -2.35 & 0.019 & 0.094 & ns\\
MRSA & Zika & 37 & 37 & -1.95 & 0.051 & 0.171 & ns\\
MRSA & Hep & 37 & 35 & -2.51 & 0.012 & 0.094 & ns\\
MRSA & Ebola & 37 & 33 & -1.19 & 0.234 & 0.469 & ns\\
nCov & Zika & 43 & 37 & 0.329 & 0.742 & 0.777 & ns\\
nCov & Hep & 43 & 35 & -0.284 & 0.777 & 0.777 & ns\\
nCov & Ebola & 43 & 33 & 1.05 & 0.296 & 0.493 & ns\\
Zika & Hep & 37 & 35 & -0.586 & 0.558 & 0.697 & ns\\
Zika & Ebola & 37 & 33 & 0.703 & 0.482 & 0.688 & ns\\
Hep & Ebola & 35 & 33 & 1.26 & 0.206 & 0.469 & ns
\\\hline
\end{tabular}} 
\end{table}

\begin{table}[h]
\caption{\revise{Dunn's test p-value summary for contact}}
\label{tab:contactDunn}
\resizebox{\linewidth}{!}{
\begin{tabular}{|l|l|l|l|l|l|l|l|}\hline
\toprule
\textbf{Group 1} & \textbf{Group 2} & \textbf{n1} & \textbf{n2} & \textbf{statistic} & \textbf{p} & \textbf{p.adj} & \textbf{Sig} \\
\toprule

HIV & MRSA & 35 & 37 & 2.88 & 0.003 & 0.059 & ns\\
HIV & nCov & 35 & 43 & 1.05 & 0.292 & 0.502 & ns\\
HIV & Zika & 35 & 37 & 0.958 & 0.338 & 0.502 & ns\\
HIV & Hep & 35 & 35 & 0.512 & 0.609 & 0.702 & ns\\
HIV & Ebola & 35 & 33 & 1.91 & 0.055 & 0.167 & ns\\
MRSA & nCov & 37 & 43 & -1.96 & 0.049 & 0.167 & ns\\
MRSA & Zika & 37 & 37 & -1.95 & 0.050 & 0.167 & ns\\
MRSA & Hep & 37 & 35 & -2.36 & 0.018 & 0.136 & ns\\
MRSA & Ebola & 37 & 33 & -0.900 & 0.368 & 0.502 & ns\\
nCov & Zika & 43 & 37 & -0.062 & 0.950 & 0.950 & ns\\
nCov & Hep & 43 & 35 & -0.516 & 0.606 & 0.702 & ns\\
nCov & Ebola & 43 & 33 & 0.969 & 0.332 & 0.502 & ns\\
Zika & Hep & 37 & 35 & -0.439 & 0.661 & 0.708 & ns\\
Zika & Ebola & 37 & 33 & 0.996 & 0.319 & 0.502 & ns\\
Hep & Ebola & 35 & 33 & 1.41 & 0.159 & 0.397 & ns
\\\hline
\end{tabular}} 
\end{table}

\begin{table}[h]
\caption{\revise{Dunn's test p-value summary for identity}}
\label{tab:identityDunn}
\resizebox{\linewidth}{!}{
\begin{tabular}{|l|l|l|l|l|l|l|l|}\hline
\toprule
\textbf{Group 1} & \textbf{Group 2} & \textbf{n1} & \textbf{n2} & \textbf{statistic} & \textbf{p} & \textbf{p.adj} & \textbf{Sig} \\
\toprule

HIV & MRSA & 35 & 37 & 2.38 & 0.017 & 0.152 & ns\\
HIV & nCov & 35 & 43 & 1.06 & 0.287 & 0.524 & ns\\
HIV & Zika & 35 & 37 & 0.454 & 0.650 & 0.740 & ns\\
HIV & Hep & 35 & 35 & 0.055 & 0.956 & 0.956 & ns\\
HIV & Ebola & 35 & 33 & 1.73 & 0.083 & 0.282 & ns\\
MRSA & nCov & 37 & 43 & -1.42 & 0.156 & 0.389 & ns\\
MRSA & Zika & 37 & 37 & -1.95 & 0.051 & 0.255 & ns\\
MRSA & Hep & 37 & 35 & -2.32 & 0.020 & 0.152 & ns\\
MRSA & Ebola & 37 & 33 & -0.590 & 0.555 & 0.694 & ns\\
nCov & Zika & 43 & 37 & -0.603 & 0.546 & 0.694 & ns\\
nCov & Hep & 43 & 35 & -1.01 & 0.314 & 0.524 & ns\\
nCov & Ebola & 43 & 33 & 0.766 & 0.444 & 0.666 & ns\\
Zika & Hep & 37 & 35 & -0.398 & 0.691 & 0.740 & ns\\
Zika & Ebola & 37 & 33 & 1.30 & 0.192 & 0.411 & ns\\
Hep & Ebola & 35 & 33 & 1.67 & 0.094 & 0.282 & ns
\\\hline
\end{tabular}} 
\end{table}
%\end{appendices}
%\input{response}

\end{document}